\documentclass[journal]{IEEEtran}
\usepackage[LGR,T1]{fontenc}
\usepackage[latin9]{inputenc}
\usepackage{array}
\usepackage{rotating}
\usepackage{float}
\usepackage{textcomp}
\usepackage{multirow}
\usepackage{amsmath}
\usepackage{amssymb}
\usepackage{graphicx}

\makeatletter

\newcommand{\lyxmathsym}[1]{\ifmmode\begingroup\def\b@ld{bold}
	\text{\ifx\math@version\b@ld\bfseries\fi#1}\endgroup\else#1\fi}

\providecommand{\tabularnewline}{\\}
\floatstyle{ruled}
\newfloat{algorithm}{tbp}{loa}
\providecommand{\algorithmname}{Algorithm}
\floatname{algorithm}{\protect\algorithmname}

\usepackage{array}
\usepackage{multirow}
\usepackage{xcolor}
\usepackage{cite}
\usepackage{algorithm}
\usepackage{algpseudocode}

\usepackage{xcolor}

\makeatother

\begin{document}
	\pagestyle{empty}
	
	\title{Processing of body-induced thermal signatures \linebreak{}
		for physical distancing and temperature screening}
	\author{Stefano Savazzi,~\IEEEmembership{member,~IEEE}, Vittorio Rampa,~\IEEEmembership{member,~IEEE},
		Leonardo Costa, Sanaz Kianoush,~\IEEEmembership{member,~IEEE}, Denis
		Tolochenko\thanks{\textcopyright  2020 IEEE.  Personal use of this material is permitted.  Permission from IEEE must be obtained for all other uses, in any current or future media, including reprinting/republishing this material for advertising or promotional purposes, creating new collective works, for resale or redistribution to servers or lists, or reuse of any copyrighted component of this work in other works.} \thanks{S. Savazzi, V. Rampa and S. Kianoush are with the Institute of Electronics,
			Computer and Telecommunication Engineering (IEIIT) of Consiglio Nazionale
			delle Ricerche (CNR), P.zza Leonardo da Vinci, 32, 20133 Milano, Italy,
			e-mail: \{stefano.savazzi, vittorio.rampa, sanaz.kianoush\}@ieiit.cnr.it.} \thanks{L. Costa and D. Tolochenko are with Cognimade S.r.l., http://www.cognimade.com,
			Segrate, Milano, Italy, e-mail: \{leonardo.costa, denis.tolochenko\}@cognimade.com.}\thanks{This work is in part funded by Regione Lombardia,
			POR-FESR 2014-2020 Innodriver framework programme. The paper has been accepted for publication in the IEEE Sensors Journal. The current arXiv contains an additional
			Appendix that describes the Envisense system configuration.}}
	
	\maketitle
	\thispagestyle{empty} 
	\begin{abstract}
		Massive and unobtrusive screening of people in public environments
		is becoming a critical task to guarantee safety in congested shared
		spaces, as well as to support early non-invasive diagnosis and response
		to disease outbreaks. Among various sensors and Internet of Things
		(IoT) technologies, thermal vision systems, based on low-cost infrared
		(IR) array sensors, allow to track thermal signatures induced by moving
		people. Unlike contact tracing applications that exploit short-range
		communications, IR-based sensing systems are passive, as they do not
		need the cooperation of the subject(s) and do not pose a threat to
		user privacy. The paper develops a signal processing framework that
		enables the joint analysis of subject mobility while automating the
		temperature screening process. The system consists of IR-based sensors
		that monitor both subject motions and health status through temperature
		measurements. Sensors are networked via wireless IoT tools and are
		deployed according to different configurations (wall- or ceiling-mounted
		setups). The system targets the joint passive localization of subjects
		by tracking their mutual distance and direction of arrival, in addition
		to the detection of anomalous body temperatures for subjects close
		to the IR sensors. Focusing on Bayesian methods, the paper also addresses
		best practices and relevant implementation challenges using on field
		measurements. The proposed framework is privacy-neutral, it can be
		employed in public and private services for healthcare, smart living
		and shared spaces scenarios without any privacy concerns. Different
		configurations are also considered targeting both industrial, smart
		space and living environments.
	\end{abstract}

	\IEEEpeerreviewmaketitle{}

\section{Introduction}
\label{sec:introduction}
\IEEEPARstart{T}{he} introduction of new sensors, digital technologies and processes \cite{transf} to monitor the mobility of the citizens is the key to address future challenges related to active health monitoring of smart spaces \cite{mag2016}, to guarantee safety and human wellness \cite{darkspaces}. In addition, the automatic verification of indoor social distance requirements, namely without the need of a human operator intervention, is expected to become critical for long-term containment of epidemic spreads, to sustain and optimize the environment, and to monitor the people behavior in shared spaces as well. The use of thermal sensors for human body sensing \cite{thermal1} is attractive in many Internet of Things (IoT) relevant scenarios, such as assisted
or smart living and industrial automation just to cite a few. In fact, these sensors are able to detect body occupancy, position and distance (both mutual and relative to walls and other objects), capturing also body surface/skin temperatures via a contactless approach.

\begin{figure}[!t]
	\center\includegraphics[scale=0.25]{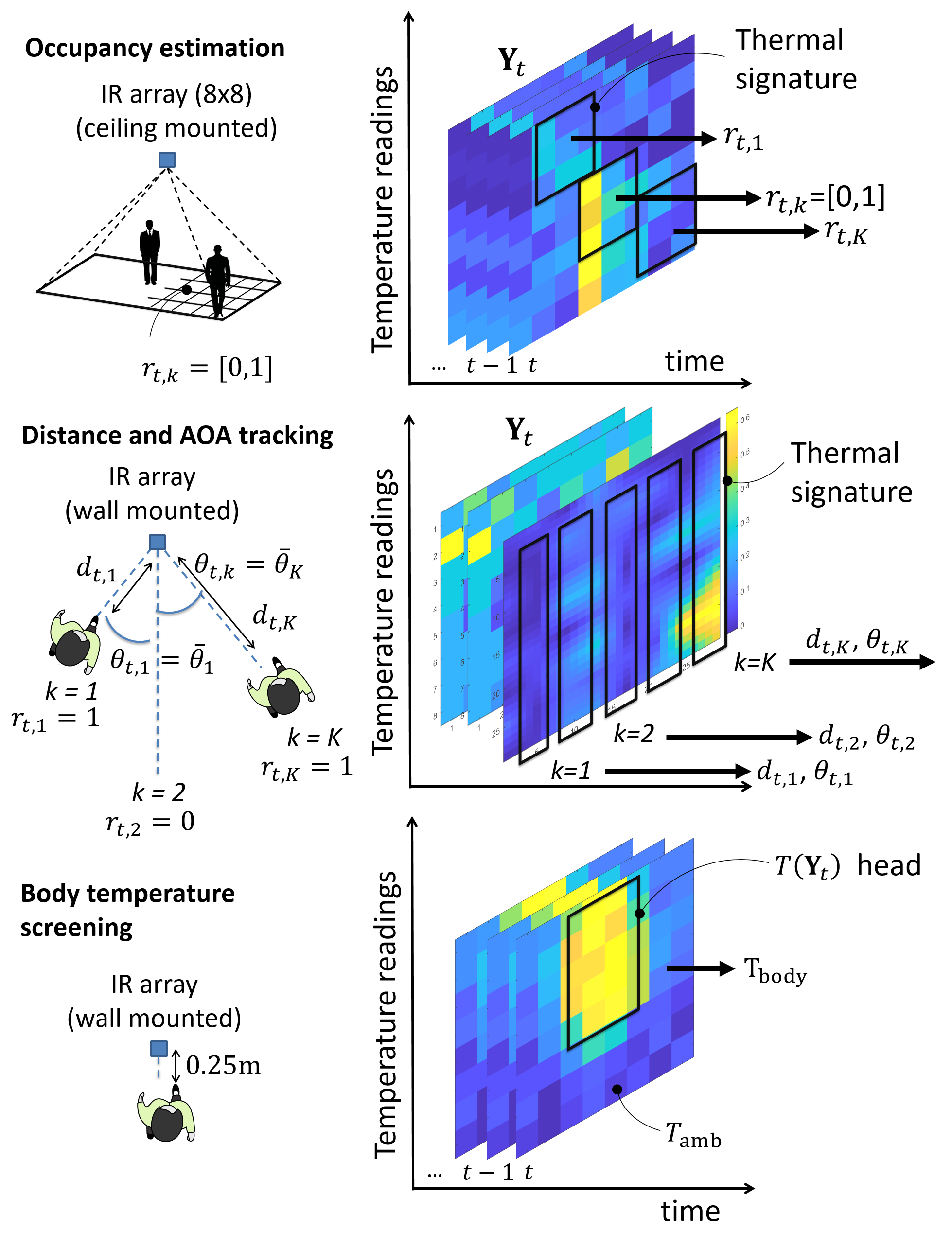} 
	\protect\caption{\label{introthermal} Bayesian models and setups for occupancy estimation,
		distance/AOA tracking and body temperature screening. Examples of thermal signature
		backlogs from EnviSense devices \cite{envisense} are also sketched.}
\end{figure}
Low-cost passive InfraRed (IR) sensors, also known as thermopiles,
are mostly used for subject detection in intrusion alarm systems and
lighting applications \cite{key-6}. Unlike emerging contact tracing
applications \cite{tracing} or other active technologies \cite{wearable},
IR-based sensing systems do not need the cooperation of the monitored
subjects. In addition, these systems enable body motion detection,
without being limited by privacy issues, since no specific person
can be recognized through the analysis of thermal frames and no electronic
records are stored. IR sensors also overcome
some known computer vision limitations \cite{thermal2-1}, since humans
thermal profile does not depend on light conditions. Finally, the
analysis of body-induced thermal signatures enable advanced applications,
\emph{i.e.} body occupancy detection, passive localization and tracking
\cite{thermal2,thermal3}.

As shown in Fig. \ref{introthermal}, the paper targets a system consisting
of multiple networked IR sensor arrays detecting the IR radiation
in the Long Wavelength InfraRed (LWIR) band. Sensors can be deployed
according to different configurations, \emph{i.e.} wall- or ceiling-mounted.
A Bayesian signal processing framework is proposed to optimize both
distancing monitoring (localization and counting) and unobtrusive
body temperature screening procedures for early diagnosis. The joint
analysis of mobility and body temperatures is validated in several
indoor operational environments exploiting real field measurements.

\subsection{Related works}

IR sensor arrays \cite{thermal2,thermal4,hvac} have typically low-cost
and are characterized by a high temporal and a low spatial resolution.
When deployed for monitoring large spaces, multiple arrays can be
also placed in selected spots to optimize coverage. Signal processing
methods usually apply to raw thermal measurements and target the estimation
of human subject(s) space occupancy in real-time, for monitoring body
positions, motion intentions such as speed, Direction/Angle Of Arrival
(DOA/AOA) \cite{key-6}, and for activity recognition \cite{activity}.
Existing techniques are mostly based on frame-based computer vision
approaches \cite{thermal1}. They consider analytics over individual
time-slices (frames) of raw temperature measurements, ranging from
K-nearest neighbor (K-NN) classifiers \cite{thermal4,hvac}, decision
trees \cite{class,key-5}, Kalman filtering \cite{key-6,eurasip},
support vector machines \cite{thermal2-1,activity} up to deep convolutional
encoder-decoder networks \cite{naser_new}. Typical challenges in
thermal vision need to face the temporal disappearance of the subjects
\cite{key-7}, due to noisy readings and external heat sources that
might be interpreted as false targets. Adaptive background subtraction
methods, thresholding or probabilistic methods \cite{temperature5}
should be therefore adopted to filter out noisy thermal sources that
are not induced by body presence. More recently, the adoption of Bayesian
tools over backlogs of thermal images has shown to prevent the problem
of human body disappearance \cite{thermal3}.

Temperature measurement through contactless devices \cite{non-invasive,investigation}
are becoming attractive as they allow the automatic screening of large
populations \cite{temperature0,tympanic,temperature}. Existing automatic
systems are capable of providing continuous real-time monitoring
at the price of a reduced user's mobility. However, contactless
instruments like thermal camera scanners, have some limitations concerning
temperature accuracy, and lack of a standardized processing interface,
tools and protocols. Several studies have been conducted to investigate
usability and reliability of thermal cameras for fever measurements
\cite{temperature5,temperature2,temperature3,temperature4,sensors1}.
Autonomy, size, and costs are generally the main issues limiting the
mass diffusion of these devices.

\subsection{Contributions}

The paper focuses on the analysis, design and verification of a unified
Bayesian framework for the real-time tracking of body-induced thermal
signatures obtained from LWIR sensor arrays arbitrarily deployed in
the monitored area. The proposed framework enables three main functions:
\emph{i) }the estimation of the body occupancy in selected spots from
which the mutual distance between subjects \cite{social} (\emph{i.e.,}
physical distancing) can be inferred; \emph{ii)} the anonymous tracking
of the relative position of the subject(s) with respect to the IR
array, namely the distance and the angle of arrival (AOA) of the targets
moving in the considered space, and \emph{iii)} the contactless body
temperature screening for subjects located nearby the sensors.

Compared to conventional frame-based methods \cite{thermal2-1,thermal2,thermal4},
the proposed framework adopts a statistical model for the extraction
of body-induced thermal signatures from noisy data, and a mobility
model to track multi-body movements and to avoid false target detection.
The approach generalizes the system proposed in \cite{thermal3} to
arbitrary sensor deployments, including wall- and ceiling-mounted
setups. It can be thus used to assess the combined use of wall- and
ceiling-mounted distributed sensors and to improve accuracy or coverage.
Body temperature screening \cite{investigation} is rooted here at
Bayesian decision theory: first, a statistical model is proposed to
relate the body surface temperature with the noisy IR sensor readings.
Next, a method for anomalous body temperature detection is developed
to account for variable air/background temperature, noisy heat sources
and small voluntary or involuntary body (\emph{e.g.}, head) movements.
It also fuses IR data with a radio-frequency (RF) low-cost radar for 
finer grained head positioning.

The paper is organized as follows: Sect. \ref{sec:System-model} introduces
the main challenges and assumptions. Focusing on occupancy estimation
and distancing, Sect. \ref{subsec:Bayesian-model-learning} targets
the problem of body-induced thermal signature modeling, while Sect.
\ref{sec:Bayesian-filtering-for} develops Bayesian filtering methods
for real-time subject tracking. Body temperature screening is described
in Sect. \ref{sec:Contact-less-body-temperature}, while Sect.
\ref{sec:A-case-study} addresses both best practices and relevant
implementation challenges, including sensor calibration, IR and radar
data fusion strategies. Finally, the performances of the proposed methods are
verified by measurements targeting both industrial and smart living
scenarios.

\begin{figure*}[!t]
	\center\includegraphics[scale=0.35]{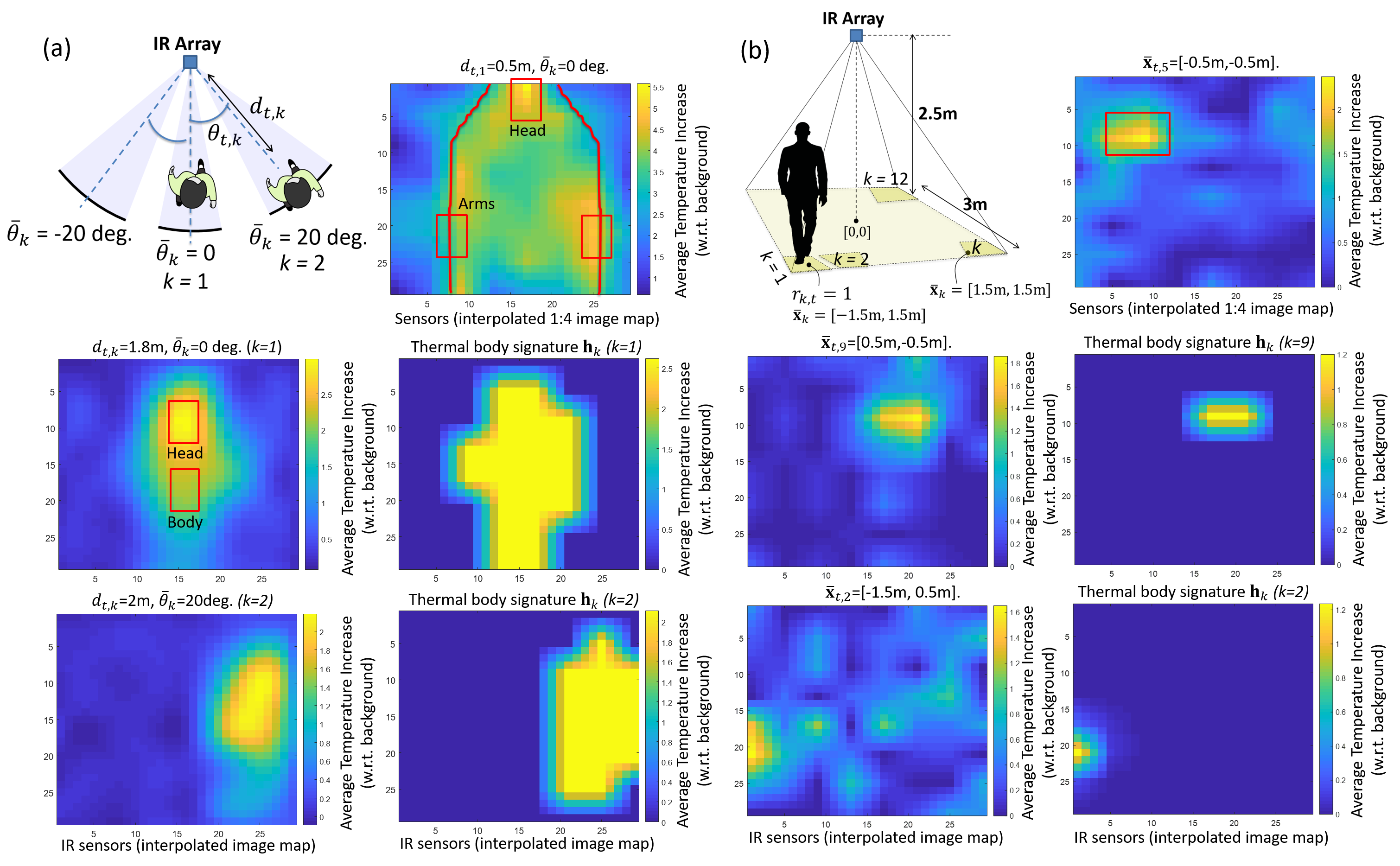} 
	\protect\caption{\label{examples} Examples of thermal signatures. (a) Wall-mounted
		IR sensors: interpolated thermal images (left) and corresponding signatures
		$\mathbf{h}_{k}=\overline{\sigma}(d_{t,k})\,\mathbf{b}_{k}$ (right)
		for different body positions (example regions $k=1,2$). (b) Ceiling-mounted
		IR sensors: thermal images (left) and signatures $\mathbf{h}_{k}=\overline{\sigma}\,\mathbf{b}_{k}$
		(right) for different body positions (example regions $k=2,9$).}
\end{figure*}

\section{System and problem definition}

\label{sec:System-model}

We introduce here the statistical model for the raw thermal data captured
by the IR sensor array. This model can serve as a general framework
for application to multi-sensor deployments and large IR arrays as
well. Sensor readings at time $t$ are collected in the vector $\mathbf{y}_{t}$
of size $M$, $\mathbf{y}_{t}=[y_{t,1},...,y_{t,M}]^{\mathrm{T}}$,
assuming $M$ thermopile elements, or detectors (in 1D linear or 2D
grid format). In the examples of Fig. \ref{introthermal}, the detector
array acquires raw thermal IR images organized as 2D frames of $8\times8$
pixels ($M=64$). For the purpose of localization, the area $\mathcal{X}$
within the Field Of View (FOV) of each IR array sensor is organized
into a grid consisting of $K\leq M$ different physical Regions Of
Interest (ROI). Each region $k$ (with $1\le k\le K$) thus covers
a 2D space $\mathcal{X}_{k}$ with $\cup_{k}\mathcal{X}_{k}=\mathcal{X}$.
As depicted in Fig. \ref{examples}(a), for wall-mounted sensors,
namely 2D sensor arrays looking inward and deployed in a vertical
plane, the $K$ ROIs represent different access areas for the subjects:
they are characterized by $\mathbf{\overline{\theta}}_{k}$, namely
the target AOA. Subject position is given in terms of distance $d$
and AOA $\theta$ relative to the IR array sensor. Similarly, for
ceiling-mounted sensors in Fig. \ref{examples}(b), namely 2D sensor
arrays looking downward and deployed in an horizontal plane, each
ROI $k$ is characterized by the relative 2D location footprint $\mathbf{\overline{x}}_{k}$
with respect to the intersection of the vertical axis passing through
the sensor array center and the horizontal plane of the floor. Considering
both setups (\emph{i.e., }wall- and ceiling-mounted sensors), the
problem we tackle is threefold:
\begin{enumerate}
	\item to estimate, for all $K$ regions, the occupancy vector $\mathbf{\mathbf{r}}_{t}=[r_{t,1},...,r_{t,k},...,r_{t,K}]^{\mathrm{T}}$
	of size $K\times1$, where $r_{t,k}\in[0,1]$ provides the binary
	information about the presence of the subject at time $t$ in the
	$k$-th region;
	\item to quantify the relative position of the subject(s) in the estimated
	region(s), \emph{i.e.} the relative 2D position of the subject $\Theta_{t,k}=[d_{t,k},\theta_{t,k}]$,
	with distance $d_{t,k}$ and AOA $\theta_{t,k}$, when crossing the
	FOV of the corresponding sensor at time $t$;
	\item to detect the body temperature of the identified subject $T_{\mathrm{body}}$
	via a contactless and non-invasive (\emph{i.e.}, camera-less) approach,
	triggered by the specific locations occupied by the body.
\end{enumerate}
The occupancy $\mathbf{r}_{t}$ and locations $\Theta_{t,k}$ can
be used to count the number of people in the area, to estimate their
positions and to monitor the mutual distance between subjects \cite{social}
as well.
The noisy temperature readings $\mathbf{y}_{t}$ depend linearly on
the occupancy pattern $\mathbf{\mathbf{r}}_{t}$ as 
\begin{equation}
\mathbf{y}_{t}=\mathbf{H}\cdot\mathbf{\mathbf{r}}_{t}+\mathbf{w}_{t}=\,\sum_{k=1}^{K}\mathbf{h}_{k}\left(\Theta_{t,k}\right)\cdot r_{t,k}+\mathbf{w}_{t},\label{eq:model}
\end{equation}
where each \emph{m}-th element $y_{t,m}$ of $\mathbf{y}_{t}$ is
thus $y_{t,m}=\negthinspace\sum_{k=1}^{K}h_{m,k}(\Theta_{t,k})\cdot r_{t,k}+w_{t,m}$.
The matrix $\mathbf{H}=\mathbf{H}(\Theta_{t,k})$, 
\begin{equation}
\mathbf{H}(\Theta_{t,k})=[\mathbf{h}_{1},...,\mathbf{h}_{k},....,\mathbf{h}_{K}]\label{eq:sign}
\end{equation}
of size $M\times K$ collects the \emph{thermal signatures} vectors
$\mathbf{h}_{k}=\mathbf{h}_{k}\left(\Theta_{t,k}\right)=[h_{1,k}\left(\Theta_{t,k}\right),...,h_{M,k}\left(\Theta_{t,k}\right)]^{\mathrm{T}}$
of size $M\times1$. Signatures describe the pattern of temperature
increases induced by a body occupying the region $k$ at position
$\Theta_{t,k}$. Clearly, in absence of the target \emph{i.e.} for
$\Theta_{t,k}=\emptyset$, it is $\mathbf{h}_{k}(\Theta_{t,k})=\mathbf{0}.$
Modeling of $\mathbf{h}_{k}$ in (\ref{eq:model}) depends on wall
or ceiling sensor layouts and it is addressed in the next sections.
The background column vector $\mathbf{w}_{t}=[w_{t,1},...,w_{t,M}]^{\mathrm{T}}$
of size $M\times1$ conveys information about detectors noise and
noisy heat-sources that are not caused by body movements but characterize
the empty space. This is modeled here as multivariate Gaussian $\mathbf{w}_{t}\sim\mathcal{N}(\boldsymbol{\mu},\mathbf{C})$
described by the average vector $\boldsymbol{\mu}=[\mu_{1},...,\mu_{M}]^{\mathrm{T}}$
and the covariance matrix $\mathbf{C}$.

\section{Body-induced thermal signature model}

\label{subsec:Bayesian-model-learning}

Learning of the body-induced thermal signatures in $\mathbf{H}$ and
the background/ambient temperature $\left\{ \boldsymbol{\mu},\mathbf{C}\right\} $
in (\ref{eq:model}) can be based on a supervised method, \emph{i.e.}
the conditional maximum likelihood $\widehat{\mathbf{H}}=\mathrm{argmax_{\mathbf{H}}}\mathrm{Pr}\left[\mathbf{y}_{t}\mid\mathbf{\mathbf{r}}_{t};\mathbf{H},\boldsymbol{\mu},\mathbf{C}\right]$
estimator. However, considering the interpolated ($1:4$) thermal
image examples of Fig. \ref{examples}, it is reasonable to assume
that the individual elements $h_{m,k}$ of matrix $\mathbf{H}$ are
intrinsically sparse \cite{thermal3}. Assuming the linear coefficients
of $\mathbf{H}$ to have a Laplace prior distribution \cite{lasso},
and Gaussian background\footnote{Ambient temperature $\boldsymbol{\mu}$ and deviations $\mathbf{C}$
	are obtained during initial sensor startup by using thermal measurements
	in the empty area.} (\ref{eq:model}), the model $\mathbf{H}$ is estimated using a Least-Squares
(LS) method:
\begin{equation}
\widehat{\mathbf{H}}=\mathrm{argmin_{\mathbf{H}}}\sum_{i=1}^{N}\left\Vert \mathbf{\widetilde{y}}_{t}^{(i)}-\mathbf{H}\cdot\mathbf{\mathbf{r}}_{t}^{(i)}\right\Vert _{\mathbf{C}}^{2}+\lambda\sum_{m=1}^{M}\sum_{k=1}^{K}\left|h_{m,k}\right|.\label{eq:model-1-1}
\end{equation}
This is based on $N$ labeled training measurements collected in the
set $\left(\mathbf{\mathbf{r}}_{t}^{(i)},\mathbf{y}_{t}^{(i)}\right)$,
namely the thermal frames $\mathbf{\widetilde{y}}_{t}^{(i)}=\mathbf{y}_{t}^{(i)}-\mathbf{\boldsymbol{\mu}}$,
$i=1,...,N$, and the corresponding true occupancy patterns $\mathbf{r}_{t}^{(i)}$.
A regularization parameter $\lambda=41$ is used for optimization,
while $\left\Vert \mathbf{y}\right\Vert _{\mathbf{C}}=\mathbf{y}^{T}\mathbf{C}^{-1}\mathbf{y}$
denotes weighting by covariance $\mathbf{C}$. Although the above
Lasso-type regularization \cite{lasso} sets matrix $\mathbf{H}$
to have a sparse representation, the solution is sensitive to training
impairments, and thus requires time-consuming calibration. In what
follows, we propose a stochastic model for thermal signatures that
is less sensitive to such impairments.

\begin{table}[tp]
	\begin{centering}
		\protect\caption{\label{table-parameters} Sensor parameters for body localization
			(thermal signature modeling for physical distancing) and body surface
			temperature measurement (temperature screening).}\vspace{-0.2cm}
		\begin{tabular}{|c|l|l|l|}
			\hline 
			\multirow{2}{*}{\begin{turn}{90}
					\textbf{Physical distancing$\negthinspace\negthinspace\negthinspace\negthinspace\negthinspace$}
			\end{turn}} & \multicolumn{2}{l|}{Wall-mounted} & $\begin{array}{l}
			\mathbf{h}_{k}(d_{t,k})=\sigma(d_{t,k})\,\mathbf{b}_{k}\textrm{ see eq. (\ref{model_thermal_sign})}\\
			\mathbf{b}_{k}:\forall k,\,\tau_{k}=\tau=0.8\,\text{°}\mathrm{C}\textrm{ (threshold)}\\
			\overline{\sigma}(d_{t,k})\textrm{ see eq. (\ref{eq:rect})}\\
			d_{min}=0.25\,\textrm{m},\,d_{max}=3.5\,\textrm{m}\\
			\sigma_{T}=1.5\,\text{°}\mathrm{\textrm{C}}\\
			\overline{\sigma}_{0}=4.5\,{^\circ}\mathrm{\textrm{C}},\,\gamma=1.1\mathrm{\,{^\circ}\textrm{C/m}}
			\end{array}$\tabularnewline
			\cline{2-4} 
			& \multicolumn{2}{l|}{Ceiling-mounted} & $\begin{array}{l}
			\mathbf{h}_{k}=\sigma\,\mathbf{b}_{k}\textrm{ see eq. (\ref{model-1})}\\
			\mathbf{b}_{k}:\forall k,\,\tau_{k}=\tau=0.4\,\text{°}\textrm{C}\textrm{ (threshold)}\\
			\overline{\sigma}=1.3\,{}^{\circ}\mathrm{C}\mathrm{\textrm{ (ceiling height \ensuremath{3\,}m)}}\\
			\sigma_{T}=0.3\,{}^{\circ}\mathrm{C}
			\end{array}$\tabularnewline
			\hline 
			\multicolumn{2}{|c|}{\textbf{Temperature screening}} & \multicolumn{2}{l|}{$\begin{array}{l}
				\alpha_{0}\negthinspace=\negthinspace0.67,\,\alpha_{1}\negthinspace=\negthinspace0.45\textrm{ (linear)}\\
				\alpha_{0}\negthinspace=\negthinspace0.66,\,\alpha_{1}\negthinspace=\negthinspace0.54,\,\alpha_{2}\negthinspace=\negthinspace-\negthinspace0.21\negthinspace\textrm{ (quad.)\negthinspace}\\
				\beta_{0}=1,\,\ensuremath{\beta_{1}=-0.09}\\
				\sigma_{\mathrm{body}}=0.4\,{^\circ}\textrm{C}\\
				\xi=-0.2\\
				Q=6\div12\textrm{ (number of samples)}
				\end{array}$}\tabularnewline
			\hline 
		\end{tabular}
		\par\end{centering}
	\medskip{}
	\vspace{-0.6cm}
\end{table}

\subsection{Signature modeling for distance and AOA estimation}

\label{subsec:Thermal-signature-modelling}

The proposed simplified model sets the individual thermal signatures
$\mathbf{h}_{k}$, namely the columns of $\mathbf{H}$, to depend
linearly on the body distances $d_{t,k}$ and have binary and sparse
representations. Signatures are thus approximated as

\begin{equation}
\mathbf{h}_{k}(d_{t,k})=\sigma(d_{t,k})\,\mathbf{b}_{k},\label{model_thermal_sign}
\end{equation}
where $\sigma(d_{t,k})$ measure the body-induced temperature increase
at distance $d_{t,k}$ and 
\begin{equation}
\mathbf{b}_{k}=\mathbf{1_{\mathit{y_{t,m}-\mu_{m}>\tau_{\mathit{k}}}}}(y_{t,m}-\mu_{m})\label{eq:simplifiedmodel-1}
\end{equation}
is a binary vector, where $\mathbf{1_{\mathit{x>\tau_{\mathit{k}}}}}(x)$
is the indicator function, namely $\mathbf{1_{\mathit{x>\tau_{\mathit{k}}}}}(x)=1$
if $x>\tau_{\mathit{k}}$ and $\mathbf{1_{\mathit{x>\tau_{\mathit{k}}}}}(x)=0$,
otherwise. The threshold $\tau_{\mathit{k}}$ is set, for each region
$k$, according to the selected layout. In Tab. \ref{table-parameters}
and Sect. \ref{subsec:Sensor-calibration}, we assume that $\forall k$
it is $\tau_{\mathit{k}}=\tau$. Coefficient $\sigma(d_{t,k})$ models
the temperature increase as a stochastic function of the body distance,
$\sigma(d_{t,k})\sim\mathcal{N}(\overline{\sigma}(d_{t,k}),\sigma_{T}^{2})$.
The average temperature increase $\overline{\sigma}(d_{t,k})$ follows
a smooth rectified linear (s-relu) model 
\begin{equation}
\overline{\sigma}(d_{t,k})=\log\left[1+\exp(\overline{\sigma}_{0}-\gamma\cdot d_{t,k})\right],\label{eq:rect}
\end{equation}
for $d_{t,k}\geq d_{min}$. Finally, deviation term $\sigma_{T}$
accounts for random, small body movements around the nominal position.
Body-induced effects decay almost linearly with the distance from
the IR array, with rate $\gamma$: in fact, when the distance increases,
not only the body is measured, but also whatever else falls within
the spot area, including the background.

\begin{figure}[!t]
	\center\includegraphics[scale=0.58]{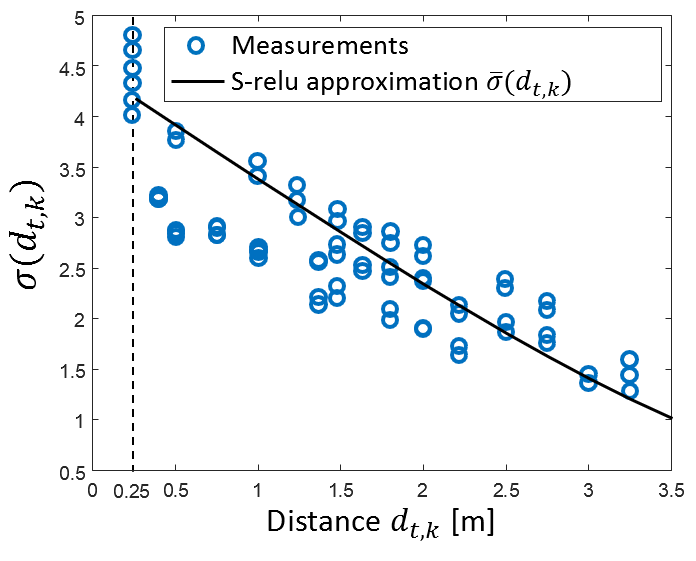} 
	\protect\caption{\label{temperatures} Model $\overline{\sigma}(d_{t,k})$ for wall
		mounted sensors and its validation from indoor measurements (s-relu
		approximation).}
\end{figure}

Using field measurements collected from $3$ subjects and two environments,
as described in Sect. \ref{sec:A-case-study}, in Fig. \ref{temperatures}
we highlight the s-relu model approximation using the parameters shown
in Tab. \ref{table-parameters} that are optimized by LS method.
Body effects converge smoothly to $\overline{\sigma}(d_{t,k})\simeq0$
when $d_{t,k}>d_{max}=3.5$ m. Examples of thermal signatures (\ref{model_thermal_sign})
for selected ROIs with AOA $\mathbf{\overline{\theta}}_{k}=\left\{ 0\lyxmathsym{\textdegree},20\lyxmathsym{\textdegree}\right\} $
are also shown in Fig. \ref{examples}(a) and compared with the averaged
temperature increases $\mathrm{\mathbb{E}}_{t}\left[\mathbf{\widetilde{y}}_{t}^{(i)}\right]$
with respect to the background.

\subsection{Signature modeling for occupancy estimation}

Considering ceiling-mounted sensors as special case, the distance
of the body from the array can be assumed as constant within the sensor
FOV. Thus, we further simplify (\ref{model_thermal_sign}) as
\begin{equation}
\mathbf{h}_{k}=\sigma\,\mathbf{b}_{k},\label{model-1}
\end{equation}
where the term $\sigma$ models the body-induced thermal increase
compared to the background. Temperature increases typically fall in
the range $\sigma=1.1^{\;\circ}\mathrm{C}\,\div\,1.4 {}^{\;\circ}\mathrm{C}$,
for $2.5\;\textrm{m}\;\div3\;\textrm{m}$ ceiling-mounted sensors, and
are consistent with the s-relu model (\ref{eq:rect}). Increase $\sigma$
is modelled here by the Gaussian probability density function $\sigma\sim\mathcal{N}(\overline{\sigma},\sigma_{T}^{2})$
where $\overline{\sigma}$ is obtained from (\ref{eq:rect}) using
the sensor height from the ground as input distance (here $2.5$ m).
For the layout of Sect. \ref{sec:A-case-study}, Fig.~\ref{examples}(b)
shows examples of the signatures $\mathbf{h}_{k}$ for a subject located
at selected positions, labelled as $\mathbf{\overline{x}}_{k}$. Signatures
are compared with the corresponding averaged temperature increases
$\mathrm{\mathbb{E}}_{t}\left[\mathbf{\widetilde{y}}_{t}^{(i)}\right]$.

\section{Bayesian filtering for localization}
\label{sec:Bayesian-filtering-for}

We estimate here, for each region
$k$, the body positions $\Theta_{t,k}$, namely the (binary) occupancy
$\Theta_{t,k}=\left[r_{t,k}\right]$, or the subject locations $\Theta_{t,k}=\left[d_{t,k},\theta_{t,k}\right],$
by considering an arbitrary sensor layout (wall or ceiling mounts).
Following the Bayesian approach, the decision on subject locations
is based on the\textit{ a-posteriori} probability $\mathrm{Pr}\left(\Theta_{t,k}\mid\mathbf{Y}_{t}\right)$
conditioned on all the observed temperature measurements $\mathbf{Y}_{t}=[\mathbf{y}_{1},...,\mathbf{y}_{t}]^{T}$
up to time $t$. By defining $\mathrm{Pr}\left(\Theta_{t,k}\mid\mathbf{Y}_{t}\right)=\Lambda_{t,k}$,
the probability terms can be evaluated iteratively as \cite{mag2016}
\begin{equation}
\Lambda_{t,k}\propto\Gamma(\mathbf{y}_{t}\mid\Theta_{t,k})\cdot\left(\sum_{h}\mathrm{Pr}(\Theta_{t,k}|\Theta_{t-1,h})\cdot\Lambda_{t-1,h}\right),\label{eq:a-post}
\end{equation}
being $\Gamma(\mathbf{y}_{t}\mid\Theta_{t,k})=\mathrm{Pr}\left(\mathbf{y}_{t}\mid\Theta_{t,k};\mathbf{h}_{k},\boldsymbol{\mu},\mathbf{C}\right)$
the \emph{conditional likelihood} and \emph{$\mathrm{Pr}\left(\Theta_{t,k}\mid\mathbf{Y}_{t-1}\right)=\sum_{h}\mathrm{Pr}(\Theta_{t,k}|\Theta_{t-1,h})\cdot\Lambda_{t-1,h}$}
the \emph{a-priori probability.} The likelihood function depends on
the background temperature ($\boldsymbol{\mu},\mathbf{C}$) and the
body-induced thermal signatures $\mathbf{h}_{k}$, defined in (\ref{model_thermal_sign})
or (\ref{model-1}) for ceiling- or wall-mounted setups, respectively,
as
\begin{equation}
\begin{array}{l}
\Gamma(\mathbf{y}_{t}\mid\Theta_{t,k})\sim\mathcal{N}\left[\overline{\sigma}\mathbf{b}_{k}+\negthinspace\mathbf{\boldsymbol{\mu}}_{k},\mathbf{C}_{k}+\sigma_{T}^{2}\mathbf{I}\right]\\
\Gamma(\mathbf{y}_{t}\mid\Theta_{t,k}=\emptyset)\sim\mathcal{N}\left[\mathbf{\boldsymbol{\mu}}_{k},\mathbf{C}_{k}\right].
\end{array}\label{eq:likelihoods}
\end{equation}
The vector $\mathbf{\boldsymbol{\mu}}_{k}=\left\{ \mu_{k,q}:\forall q,\text{ \ensuremath{b_{k,q}>0}}\right\} $
considers only the elements from $\mathbf{\boldsymbol{\mu}}$ corresponding
to the non-zero indexes of the binary vector $\mathbf{b}_{k}$ (\ref{eq:simplifiedmodel-1}),
while $\mathbf{\mathbf{C}}_{k}=\mathbb{E}_{t}\left[(\mathbf{w}_{t,k}-\boldsymbol{\mu}_{k})(\mathbf{w}_{t,k}-\boldsymbol{\mu}_{k})^{\mathrm{T}}\right]$
is defined similarly using $\mathbf{w}_{t,k}=\left\{ w_{t,q}:\forall q,\text{ \ensuremath{b_{k,q}>0}}\right\} $.
The \emph{transition probability} $\mathrm{Pr}(\Theta_{t,k}|\Theta_{t-1,h})$
is computed according to the selected layout-specific mobility model \cite{random-walk}.

\begin{algorithm}[t]
	\caption{Distancing and AOA tracking}
	
	\label{distAOA}\begin{algorithmic}[1]
		
		\Procedure{tracking}{$[\mathbf{b}_{k},\overline{\theta}_{k}]_{k=1}^{K},\overline{\sigma}(\cdot),\sigma_{T}^{2},\boldsymbol{\mu},\mathbf{C},[d]_{d=d_{min}}^{d_{max}}$}
		
		\State initialize multivariate Gaussian functions:
		
		\State $\mathrm{\mathcal{K}}(\mathbf{C})\leftarrow\left(2\pi\right)^{-\frac{M}{2}}\mathrm{det}(\mathbf{C})^{-\frac{1}{2}}$
		
		\State $\mathcal{N}\left[\mathbf{y\mathrm{;}}\mathbf{x},\mathbf{C}\right]\leftarrow\mathrm{\mathcal{K}}(\mathbf{C})\exp\left[-\frac{1}{2}(\mathbf{y-x})^{\mathrm{T}}\mathbf{C}^{-1}(\mathbf{y-x})\right]$
		
		\State $\forall d:\,\mathrm{Pr}(d|h)\leftarrow\mathcal{N}\left[d\mathrm{;}h,\mathrm{W}\right]$\Comment{random
			walk ($0.5$ m/s)}
		
		\For{each round $t$}\Comment{Main loop}
		
		\State $\forall k,d$: $\Lambda_{t,k}(d)=\frac{1}{d_{\mathrm{max}}-d_{\mathrm{min}}}$\Comment{initialize
			$\Lambda_{t,k}(d)$}
		
		\For{$k\leftarrow1,K$}\Comment{Occupancy $\widehat{r}_{t,k}$}
		
		\State$\Gamma\left(\mathbf{y}_{t}\mid r_{t,k}=1\right)\leftarrow0$
		
		\State$\Gamma\left(\mathbf{y}_{t}\mid r_{t,k}=0\right)\gets\mathcal{N}\left[\mathbf{y}_{t};\mathbf{\boldsymbol{\mu}}_{k},\mathbf{C}_{k}\right]$
		
		\For{$d\leftarrow d_{min},d_{max}$,$\Delta d$}
		
		\State$\mathbf{x}_{k}\leftarrow\overline{\sigma}(d)\mathbf{b}_{k}+\negthinspace\mathbf{\boldsymbol{\mu}}_{k}$
		
		\State$\Gamma\left(\mathbf{y}_{t}\mid d\right)\negthinspace\gets\negthinspace\mathcal{N}\left[\mathbf{y}_{t};\mathbf{x}_{k},\mathbf{C}_{k}+\sigma_{T}^{2}\mathbf{I}\right]$\Comment{$\negthinspace$(\ref{eq:likelihoods})}
		
		\State$\Gamma\negthinspace\left(\negthinspace\mathbf{y}_{t}\negthinspace\mid\negthinspace r_{t,k}\negthinspace=\negthinspace1\right)\negthinspace\negthinspace\gets\negthinspace\negthinspace\Gamma\negthinspace\left(\negthinspace\mathbf{y}_{t}\negthinspace\mid\negthinspace r_{t,k}\negthinspace=\negthinspace1\right)\negthinspace+\negthinspace\frac{\Gamma\left(\mathbf{y}_{t}\mid d\right)}{d_{\mathrm{max}}-d_{\mathrm{min}}}$
		
		\EndFor
		
		\State$\widehat{r}_{t,k}\gets\arg\max_{r_{t,k}\in[0,1]}\Gamma\left(\mathbf{y}_{t}\mid r_{t,k}\right)$
		
		\EndFor
		
		\For{all $k$ s.t. $\widehat{r}_{t,k}=1$}\Comment{AOA:$\widehat{\theta}_{t,k}$,
			dist.: $\widehat{d}_{t,k}$}
		
		\State$\widehat{\theta}_{t,k}\leftarrow\overline{\theta}_{k}$\Comment{AOA}
		
		\For{$d\leftarrow d_{min},d_{max},\Delta d$}\Comment{a-priori}
		
		\If{ $\widehat{r}_{t-1,k}=1$}\Comment{subject at $t-1$}
		
		\State $\Lambda_{t,k}(d)\negthinspace\negthinspace\leftarrow\negthinspace\negthinspace\sum_{h=d_{min}}^{d_{max}}\negthinspace\mathrm{Pr}(d|h)\times\Lambda_{t-1,k}(h)$
		
		\EndIf
		
		\State $\Lambda_{t,k}(d)\leftarrow$$\Gamma\left(\mathbf{y}_{t}\mid d\right)\Lambda_{t,k}(d)$\Comment{a-posteriori}
		
		\EndFor
		
		\State$\widehat{d}_{t,k}\gets\arg\max_{d}\Lambda_{t,k}(d)$\Comment{distancing}
		
		\EndFor
		
		\EndFor
		
		\EndProcedure
		
	\end{algorithmic}
\end{algorithm}

\subsection{Distancing and AOA monitoring}

For monitoring body distances and movements $\Theta_{t,k}=\left[d_{t,k},\theta_{t,k}\right]$,
we resort to the pseudo-code detailed in the Algorithm \ref{distAOA}
and described in the following for wall-mounted IR sensors. As depicted
in Fig. \ref{examples}(a), each ROI $k$ corresponds to a different
access area, with AOA $\mathbf{\overline{\theta}}_{k}$ (\emph{i.e.,}
azimuth) and unknown distance $d_{t,k}$, but in the range $\left[d_{\mathrm{min}},d_{\mathrm{max}}\right]$.
We first detect the access area $\hat{r}_{t,k}$, by solving
the occupancy problem. Then, both the AOA $\widehat{\theta}_{t,k}=\overline{\theta}_{k}(\hat{r}_{t,k})$
and the distance estimate $\widehat{\Theta}_{t,k}=\widehat{d}_{t,k}$
(distancing) are considered. 
Occupancy detection is based on the maximum likelihood (ML) algorithm
\begin{equation}
\widehat{r}_{t,k}=\arg\max_{r_{t,k}\in[0,1]}\mathrm{Pr}\left(\mathbf{y}_{t}\mid r_{t,k}\right)\label{eq:ML-1}
\end{equation}
with $\mathrm{Pr}\left(\mathbf{y}_{t}\mid r_{t,k}=0\right)\sim\mathcal{N}\left[\mathbf{\boldsymbol{\mu}}_{k},\mathbf{C}_{k}\right]$
and 
\begin{equation}
\mathrm{Pr}\left(\mathbf{y}_{t}\mid r_{t,k}=1\right)=\int_{d_{\mathrm{min}}}^{d_{\mathrm{max}}}\mathrm{Pr}(d_{t,k}=x)\,\Gamma(\mathbf{y}_{t}\mid\Theta_{t,k})\mathrm{d}x,\label{eq:simplified-1}
\end{equation}
where $\text{\ensuremath{\Theta_{t,k}}= \ensuremath{\left[\ensuremath{d_{t,k}},\theta_{t,k}=\overline{\theta}_{k}\right]}}$
and $d_{\mathrm{min}}\leq d_{t,k}\leq d_{\mathrm{max}}$. The integral
function in (\ref{eq:simplified-1}) is implemented as a finite sum
with interval $\Delta d=0.25$ m while $\mathrm{Pr}(d_{t,k}=x)$ is
assumed to be uniformly distributed as $\mathrm{Pr}(d_{t,k}=x)=\tfrac{1}{d_{\mathrm{max}}-d_{\mathrm{min}}}$.

Distance estimation is computed as
\begin{equation}
\hat{d}_{t,k}=\arg\max_{\widehat{r}_{t,k}=1,\text{ }d_{\mathrm{min}}\leq d_{t,k}\leq d_{\mathrm{max}}}\mathrm{Pr}\left(d_{t,k}\mid\mathbf{Y}_{t}\right)\label{eq:maxdistance}
\end{equation}
for all the occupied access areas for which $\widehat{r}_{t,k}=1,$$\forall k$:
it is obtained via Bayesian filtering (\ref{eq:a-post}) now with
$\Lambda_{t,k}=\mathrm{Pr}\left(d_{t,k}\mid\mathbf{Y}_{t}\right)$
and $\text{\ensuremath{\Theta_{t,k}}= \ensuremath{\left[\ensuremath{d_{t,k}}\right]}}$
where the conditional probability $\mathrm{\mathrm{Pr}(\Theta_{\mathit{t,k}}|\Theta_{\mathit{t\mathrm{-1},h}})}$
is given by a 1D random walk \cite{random-walk}. The subject(s) in
the area $k$ may thus move backward or forward along the range segment
$\left[d_{\mathrm{min}},d_{\mathrm{max}}\right]$ with AOA $\widehat{\theta}_{t,k}=\overline{\theta}_{k}$.

\subsection{Subject counting}

Subject counting is drawn from the ceiling-mounted IR sensors. In
fact, they can be more easily deployed in crowded areas, compared
with wall-mounted setups. Counting requires first to detect the presence
$\hat{r}_{t,k}$ of the subjects in the $K$ ROIs, then to identify
the number of occupied spots $S_{t}=\sum_{k=1}^{K}\hat{r}_{t,k}$
at time $t$. The goal is to detect critical distancing situations
(\emph{i.e.}, crowded areas).
We adopt here the Bayesian filtering (\ref{eq:a-post}) and
track the (binary) occupancy vector $\widehat{\mathbf{r}}_{t}=[\hat{r}_{t,k}]_{k=1}^{K}$
according to the Maximum A Posteriori (MAP) estimate

\begin{equation}
\widehat{\mathbf{r}}_{t}=\arg\max_{r_{t,k}\in[0,1],\forall k=1,...K}\underset{\mathrm{Pr}(\mathbf{r}_{t}\mid\mathbf{Y}_{t})}{\underbrace{\prod_{k=1}^{K}\mathrm{Pr}\left(r_{t,k}\mid\mathbf{Y}_{t}\right)}},\label{eq:map_estimate}
\end{equation}
where $\mathrm{Pr}\left(r_{t,k}\mid\mathbf{Y}_{t}\right)\propto\mathrm{Pr}\left(\mathbf{y}_{t}\mid r_{t,k}\right)\mathrm{Pr}\left(r_{t,k}\mid\mathbf{Y}_{t-1}\right)$.
Notice that the \textit{a-priori} probabilities $\mathrm{Pr}\left(r_{t,k}\mid\mathbf{Y}_{t-1}\right)$
are now binary and, as in (\ref{eq:a-post}), they can be updated
iteratively as
\begin{equation}
\mathrm{Pr}\left(r_{t,k}\mid\mathbf{Y}_{t-1}\right)=\sum\mathrm{Pr}(r_{t,k}\mid\mathbf{r}_{t-1})\mathrm{\;Pr}(\mathbf{r}_{t-1}\mid\mathbf{Y}_{t-1}),\label{eq:ITER}
\end{equation}
by summing over all (binary) combinations in $\mathbf{r}_{t-1}$.
The transition probability $\mathrm{Pr}(r_{t,k}\mid\mathbf{r}_{t-1})$
is drawn from the motion model as shown in the eq. 8 of the reference
\cite{thermal3}.

\begin{figure}[!t]
	\center\includegraphics[scale=0.5]{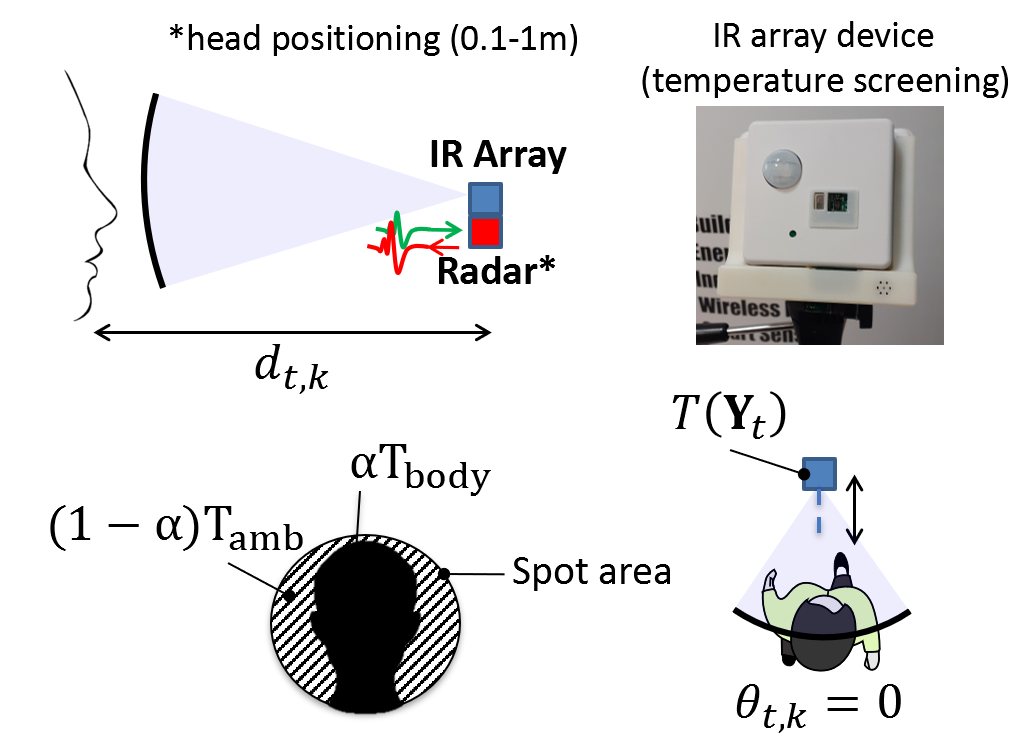} 
	\protect\caption{\label{temperature_screening} Contactless body screening setup and
		main parameters.}
\end{figure}

\section{Temperature screening}

\label{sec:Contact-less-body-temperature}

The analysis proposed in this section targets the automation of the
temperature screening process, namely the problem of massive and unobtrusive
identification of anomalous body temperatures in public environments,
where keeping user privacy is critical. The temperature measurement
is automatically triggered when the user is located near the IR sensor
array. As introduced in Sect. \ref{subsec:Thermal-signature-modelling},
the temperature readings obtained from the IR sensor do not reveal
the precise body temperature, as measurements are influenced by the
(time-varying) background and the ambient (\emph{i.e.}, air) temperature
$T_{\mathrm{amb}}$. In addition, body-induced effects also decay
almost linearly with the distance of the body from the IR sensors.

Compared with state-of-the-art low-cost IR devices \cite{temperature2,size-of-source},
the proposed screening method draws from a statistical model and Bayesian
decision theory: the user is allowed to move during the screening
process while its position is estimated (continuously) using the Bayesian
framework proposed in Sect. \ref{sec:Bayesian-filtering-for}. The
screening process gives the probability that the estimated surface
body temperature $T_{\mathrm{body}}$ rises above a given threshold
$T_{\mathrm{max}}$ ($37.5$ °C). This indicator can be used as initial
fever screening, to select the subject(s) that require a more careful
treatment (\emph{e}.\emph{g}., using manual thermometers or higher-precision
devices). Screening precision and recall figures decay with the body
distance from the IR array and positioning accuracy: this is quantified
in Sect. \ref{sec:A-case-study}, experimentally.

\begin{figure}[!t]
	\center\includegraphics[scale=0.52]{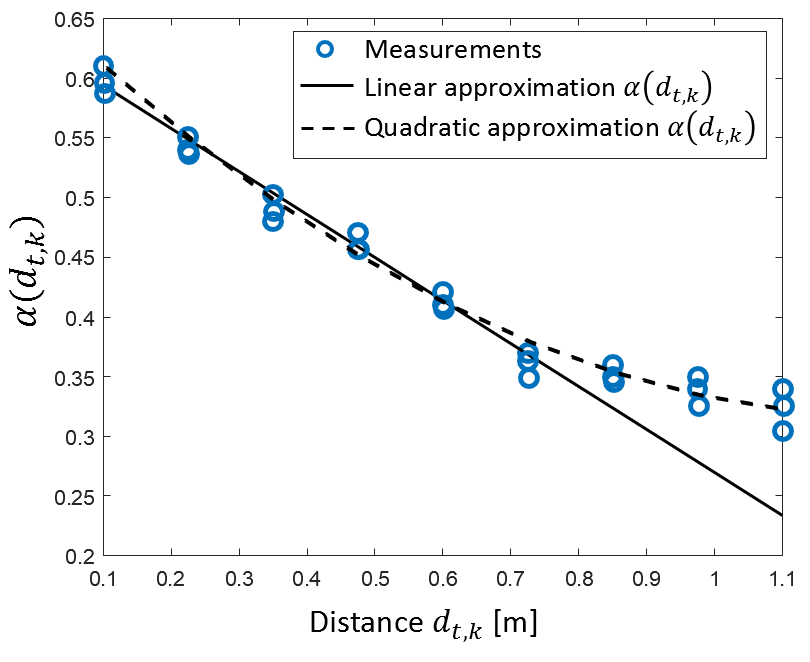} 
	\protect\caption{\label{temperature_screening_model} Linear and quadratic models $\alpha(\Theta_{t,k})$
		for temperature screening and varying body distance $\Theta_{t,k}=[d_{t,k}]$.}
\end{figure}

\subsection{Body surface temperature modeling}

Body surface temperature measurement is obtained from sensor readings
$\mathbf{y}_{t}$ through the Gaussian function $T(\mathbf{y}_{t})\sim\mathcal{N}(\overline{T}(\mathbf{y}_{t}),\sigma_{\mathrm{body}}^{2})$.
The average absolute temperature $\overline{T}(\mathbf{y}_{t})$ is
\begin{equation}
\overline{T}(\mathbf{y}_{t})=\frac{1}{Q}\sum_{\tau\in[t-Q+1,\mbox{ }t]}\max_{m=1,...,M}\left[y_{\tau,m}-\mu_{m}\right]+\mu_{m},\label{eq:maxt}
\end{equation}
over the inspection interval $[t-Q+1,\mbox{ }t]$ where $\mu_{m}$
is the \emph{m}-th thermopile background average $\mu_{m}=\mathrm{\mathbb{E}}_{t}\left[y_{t,m}\right]$.
The term $\sigma_{\mathrm{body}}^{2}$ accounts for time-varying readings
due to random, voluntary or involuntary head/body movements. The (true)
body temperature $T_{\mathrm{body}}$ can be obtained as
the solution of 
\begin{equation}
\beta\overline{T}(\mathbf{y}_{t})=\alpha\,T_{\mathrm{body}}+\left[1-\alpha\right]T_{\mathrm{amb}},\label{eq:linear}
\end{equation}
where $T_{\mathrm{amb}}$ corresponds to the ambient temperature.
Coefficient $\beta$ models the change of the temperature measurement
with $T_{\mathrm{amb}}$, or ambient-to-skin difference \cite{temperature4}.
In addition, it corrects the absolute temperature measurement error
($\pm2.5$ °C typical) observed during initial sensor startup: 
\begin{equation}
\beta=\beta_{0}\left(1+\beta_{1}\frac{T_{\mathrm{amb}}-T_{min}}{T_{min}}\right),\label{eq:beta}
\end{equation}
with $T_{min}=20{^\circ}\textrm{C}.$ Ambient temperature $T_{\mathrm{amb}}$
can be obtained directly from the background $\boldsymbol{\mu}$ in
(\ref{eq:model}) as $T_{\mathrm{amb}}=\frac{1}{M}\sum_{m=1}^{M}\mu_{m}$
or drawn from other sensors. Considering a body at distance $d_{t,k}$
and AOA $\theta_{t,k}$, the coefficient $\alpha=\alpha(d_{t,k})$,
with $0<\alpha<1$, models the fraction of the sensor spot area size
that is occupied by the body (Fig. \ref{temperature_screening}).
According to \cite{size-of-source}, it linearly depends on the body
distance $d_{t,k}$ as

\begin{equation}
\alpha(d_{t,k})=\alpha_{0}-\alpha_{1}d_{t,k}.\label{eq:alf}
\end{equation}
However, the LS model fitting depicted in Fig. \ref{temperature_screening_model}
shows that the linear approximation (\ref{eq:alf}) is effective as
far as the body distance is lower than $0.75$ m. On the contrary,
the quadratic approximation $\alpha(d_{t,k})=\alpha_{0}-\alpha_{1}d_{t,k}-\alpha_{2}d_{t,k}^{2}$,
superimposed for comparison on the same figure, gives improvements
for larger distances. The experiments in Sect. \ref{sec:A-case-study}
further reveal that the model (\ref{eq:maxt}) gives good results
as far as $20{^\circ}\textrm{C}\leq T_{\mathrm{amb}}\leq28{^\circ}\textrm{C}.$
Notice that for more general applicability, the relative humidity
of the air should be considered as well \cite{temperature}. All optimized
parameters are summarized in Tab. \ref{table-parameters} assuming
a FOV of the IR array equal to $60\lyxmathsym{\textdegree}$.

\subsection{Temperature screening: detection and alerts}

Temperature screening discriminates an anomalous body temperature
observation ($T_{\mathrm{body}}>T_{\mathrm{max}}$), described here
as state $\mathrm{F}_{1}$, from a safe condition ($T_{\mathrm{body}}<T_{\mathrm{max}}$),
indicated as $\mathrm{F}_{0}$. Decision at time $t$ is made by collecting
$Q$ samples $\mathbf{Y}_{t,Q}=[\mathbf{y}_{t-Q+1},...,\mathbf{y}_{t}]^{T}$
within the inspection interval $[t-Q+1,\mbox{ }t]$. The log-likelihood
ratio (LLR) function
\begin{equation}
\mathrm{LLR}(\mathbf{y}_{\mathit{t}})=\mathrm{log}\left[\frac{\Pr\left(\mathbf{y}_{t}|\mathrm{F}_{1}\right)}{\Pr\left(\mathbf{y}_{t}|\mathrm{F}_{0}\right)}\right]\label{eq:llr}
\end{equation}
is used as decision metric: for states $\mathrm{F}_{1}$ and $\mathrm{F}_{0}$,
the likelihoods are $\mathrm{Pr}\left(\mathbf{y}_{t}|\mathrm{F}_{1};\Theta_{t,k},T_{\mathrm{amb}}\right)$
and $\mathrm{Pr}\left(\mathbf{y}_{t}|\mathrm{F}_{0};\Theta_{t,k},T_{\mathrm{amb}}\right),$
respectively. In particular, using (\ref{eq:maxt}) and (\ref{eq:linear}),
it is 
\begin{equation}
\Pr\left(\mathbf{y}_{t}|\mathrm{F}_{1}\right)\negthinspace=\negthinspace\Pr\left[T(\mathbf{y}_{t})\negthinspace>\negthinspace\alpha T_{\mathrm{max}}\negthinspace+\negthinspace(1\negthinspace-\negthinspace\alpha)T_{\mathrm{amb}}\right]\negthinspace\frac{\Gamma(\mathbf{y}_{t}\mid\Theta_{t,k})}{\mathrm{\Pr\left(\mathrm{F}_{1}\right)}}\label{eq1-IV}
\end{equation}
being $\Gamma(\mathbf{y}_{t}\mid\Theta_{t,k})=\mathrm{Pr}\left(\mathbf{y}_{t}\mid r_{t,k},\mathbf{h}_{k}\right)$
the conditional likelihood (\ref{eq:likelihoods}) and $\Pr\left[T(\mathbf{y}_{t})\negthinspace>\negthinspace y\right]=\mathcal{Q}\left(\frac{y-\overline{T}(\mathbf{y}_{t})}{\sigma_{\mathrm{body}}}\right)$
with $\mathcal{Q}\left(\cdot\right)$ the Q-function \cite{Q-function}.
We also assume that $\mathrm{F}_{0}$ and $\mathrm{F}_{1}$ have the
same a-priori probabilities $\Pr\left(\mathrm{F}_{0}\right)=\Pr\left(\mathrm{F}_{1}\right)=1/2$.
$\mathrm{Pr}\left(\mathbf{y}_{t}|\mathrm{F}_{0};\Theta_{t,k},T_{\mathrm{amb}}\right)$
is defined similarly and thus omitted.

Being $[t-Q+1,\mbox{ }t]$ the inspection interval, the detector
implements a majority voting policy: it votes for state
$\mathrm{F}_{1}$ iff 
\begin{equation}
\sum_{\tau\in[t-Q+1,\mbox{ }t]}\mathbf{1}_{\mathrm{LLR}(\mathbf{y}_{\tau})\geq\xi}>\sum_{\tau\in[t-Q+1,\mbox{ }t]}\mathbf{1}_{\mathrm{LLR}(\mathbf{y}_{\tau})<\xi}\,.\label{eq3n-IV}
\end{equation}
The detector also outputs a soft indicator, namely the sample probability
$\frac{1}{Q}\sum_{\tau\in[t-Q+1,\mbox{ }t]}\mathbf{1}_{\mathrm{LLR}(\mathbf{y}_{\tau})\geq\xi}$,
that quantifies the severity of the alert defined in the $\mathrm{[0,1]}$
interval. The optimal threshold value $\xi$ is obtained from a measurement
campaign and it is optimized in Sect. \ref{sec:A-case-study} targeting
the maximization of the true positive rates (\emph{i.e.}, or recall
figure), as critical for preliminary screening and diagnosis operations.
\begin{figure}[!t]
	\center\includegraphics[scale=0.42]{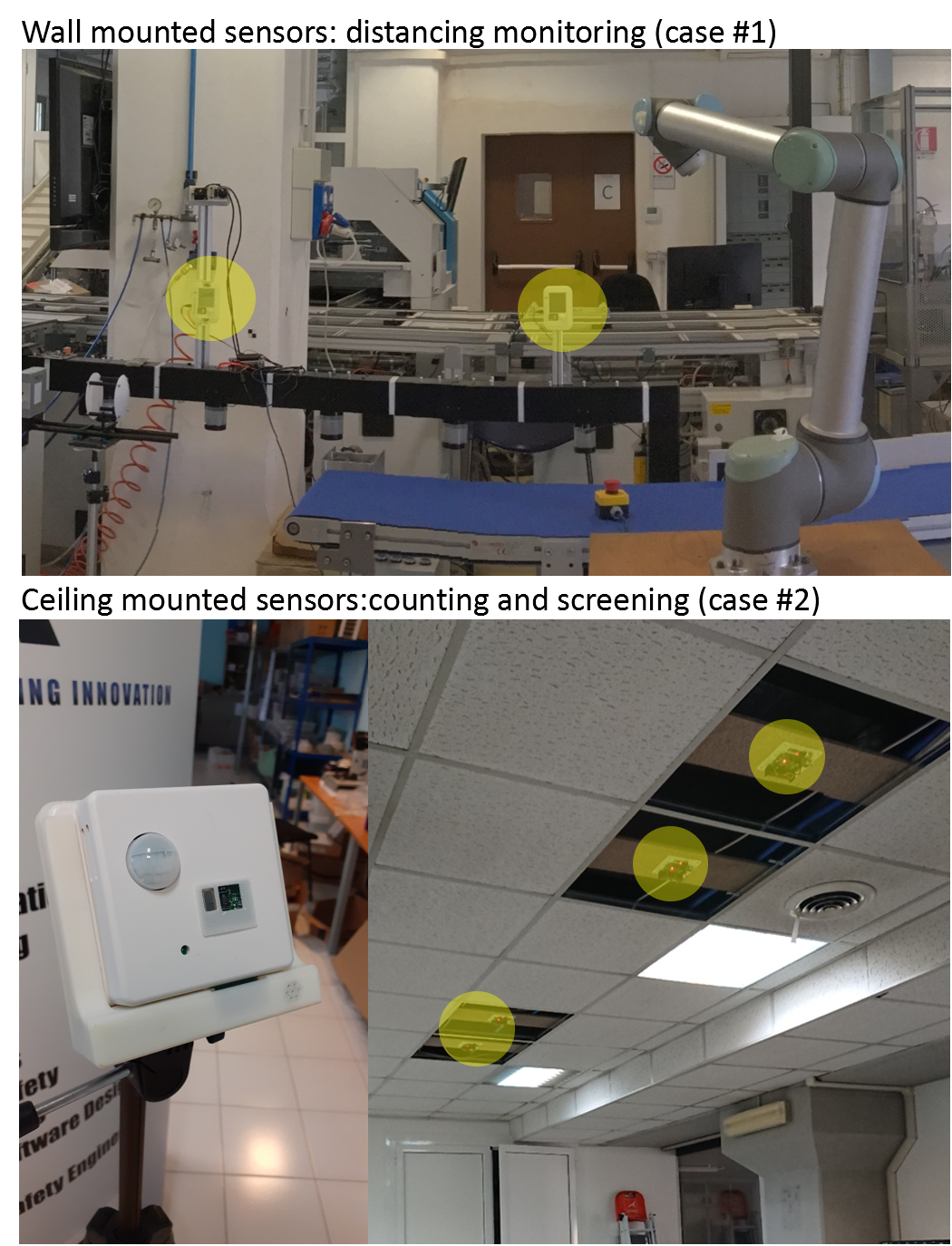} 
	\protect\caption{\label{layout} From top to bottom: wall-mounted setup (case \#1)
		and ceiling-mounted (case \#2) for different deployments inside the
		CNR-STIIMA (http://www.stiima.cnr.it/en) test plant.}
\end{figure}
\begin{figure}[!t]
	\center\includegraphics[scale=0.45]{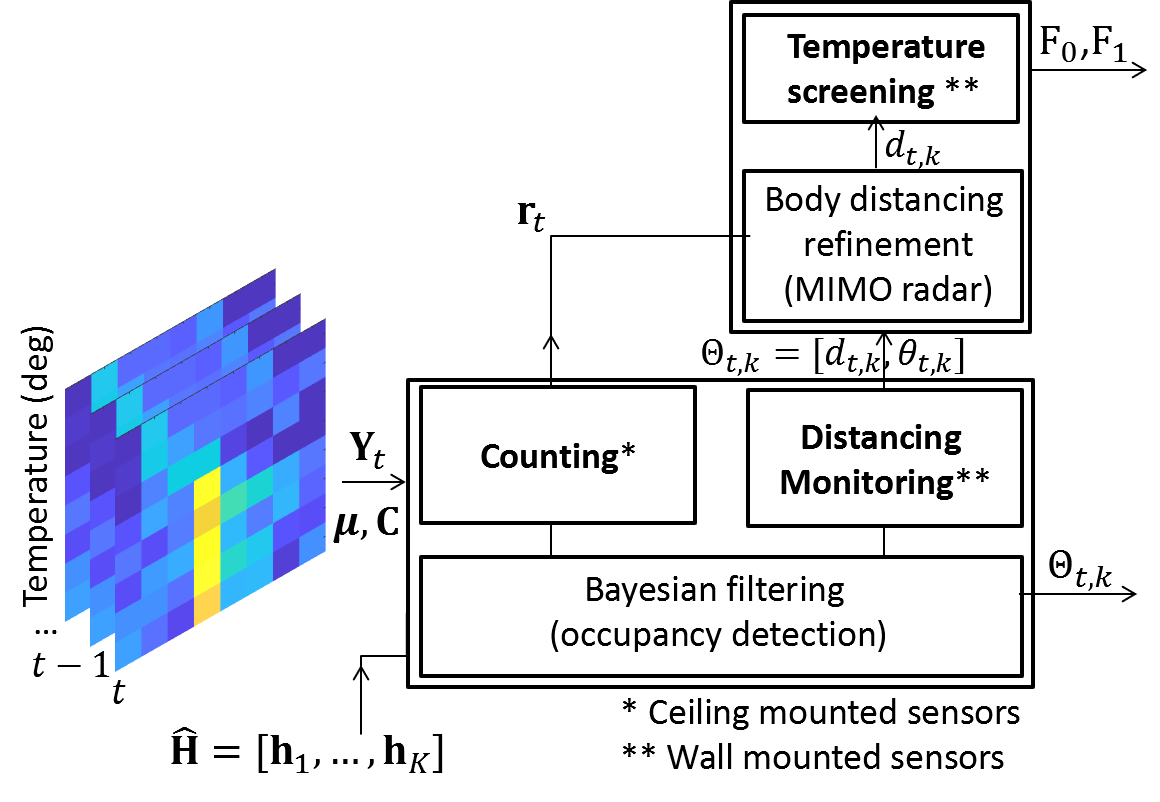} 
	\protect\caption{\label{block_diag} Block diagram for joint occupancy detection, distance
		monitoring and contactless body temperature screening.}
\end{figure}

\section{Experiments in workspace environments\label{sec:A-case-study}}

The experimental validation scenarios exploit IoT devices \cite{envisense} equipped
with thermopile sensor arrays each consisting of $M=64$ IR detectors and FOV of
$60\lyxmathsym{\textdegree}$. They are sensible in the $8-13$~$\mu$m
LWIR infrared band, with a noise equivalent temperature difference
of~$\pm0.08$ °C @ $1$ Hz at room temperature ($T_{\mathrm{amb}}=20$ °C).
In the scenario $\#1$ on top of Fig.~\ref{layout}, two wall-mounted
devices are installed to monitor the subjects access areas and their
distances from a manipulator machine \cite{fall}, for safety purposes.
In the scenario $\#2$ at the bottom of Fig.~\ref{layout}, four
devices are mounted on different ceiling panels. They monitor
a corridor and count the number of moving subjects, thus analyzing
motion patterns in real-time to generate an alert when crowded areas
are detected. A wall-mounted device is also installed nearby the entrance
for temperature screening while the subject is moving towards the
corridor.

The proposed processing framework for \emph{joint} localization \emph{and}
body temperature screening is summarized in the block diagram of Fig.
\ref{block_diag}. The IR sensor array readings $\mathbf{Y}_{t}$
are processed continuously and independently for counting, mutual
distance monitoring (ceiling-mounted sensors), localization and temperature
screening (wall-mounted). 

In Sect. \ref{subsec:Sensor-calibration},
we first detail the sensor calibration procedures and the necessary
initialization stages. Optimized model parameters are summarized in
Tab. \ref{table-parameters}. Localization is addressed in Sect.
\ref{subsec:Localization-and-distancing} by analyzing the subject
positioning accuracy (distance and AOA) and the ability to identify
crowded areas for social distancing monitoring as well. 

Finally, temperature screening in Sect. \ref{subsec:Body-temperature-detection} is 
corroborated by an experimental study involving both real and simulated body surfaces.
The impact of the inspection interval $[t-Q+1,\mbox{ }t]$ and the
subject positioning errors on temperature estimation is also discussed.
All proposed Bayesian methods (Sect. \ref{sec:Bayesian-filtering-for}
and \ref{sec:Contact-less-body-temperature}) are implemented on a
low-power System on Chip (SoC) exploiting a $1.5$ GHz quad core ARM
Cortex-A72 processor with 4 GB internal RAM.

\begin{table}[tp]
	\begin{centering} \protect\caption{\label{tab_distance_AOA} Distance and AOA estimation RMSE vs. $\overline{\sigma}_{0}$.}\vspace{-0.2cm}
		\begin{tabular}{|c|c|c|c|c|c|c|}
			\cline{2-7} 
			\multicolumn{1}{c|}{} & \multicolumn{2}{c|}{RMSE} & \multicolumn{2}{c|}{RMSE} & \multicolumn{2}{c|}{RMSE}\tabularnewline
			\cline{2-7} 
			\multicolumn{1}{c|}{} & \multicolumn{2}{c|}{$\overline{\sigma}_{0}=4.5{^\circ}\mathrm{\textrm{C}}$ (Tab. \ref{table-parameters})} & \multicolumn{2}{c|}{$\overline{\sigma}_{0}=5.5{^\circ}\mathrm{\textrm{C}}$} & \multicolumn{2}{c|}{$\overline{\sigma}_{0}=3.5{^\circ}\mathrm{\textrm{C}}$}\tabularnewline
			\hline 
			$d_{t,k}$ & $\widehat{\theta}_{t,k}$ & $\widehat{d}_{t,k}$ & $\widehat{\theta}_{t,k}$ & $\widehat{d}_{t,k}$ & $\widehat{\theta}_{t,k}$ & $\widehat{d}_{t,k}$\tabularnewline
			\hline 
			\multirow{1}{*}{$0.5$ m} & $8.8^{\circ}$ & $0.32$ m & $7.7{^\circ}$ & $0.40$ m & $14.8{^\circ}$ & $0.30$ m\tabularnewline
			\hline 
			\multirow{1}{*}{$1.0$ m} & $4.9^{\circ}$ & $0.28$ m & $4.8{^\circ}$ & $0.45$ m & $10.7{^\circ}$ & $0.28$ m\tabularnewline
			\hline 
			$1.5$ m & $4.2^{\circ}$ & $0.44$ m & $3.7{^\circ}$ & $0.54$ m & $9.2{^\circ}$ & $0.34$ m\tabularnewline
			\hline 
			$2.0$ m & $4.5^{\circ}$ & $0.52$ m & $3.5{^\circ}$ & $0.52$ m & $10.5{^\circ}$ & $0.44$ m\tabularnewline
			\hline 
			$2.5$ m & $4.2^{\circ}$ & $0.67$ m & $3.1{^\circ}$ & $0.70$ m & $10.2{^\circ}$ & $0.51$ m\tabularnewline
			\hline 
		\end{tabular}
		\par\end{centering}
	\medskip{}
\end{table}
\begin{figure}[!t]
	\center\includegraphics[scale=0.36]{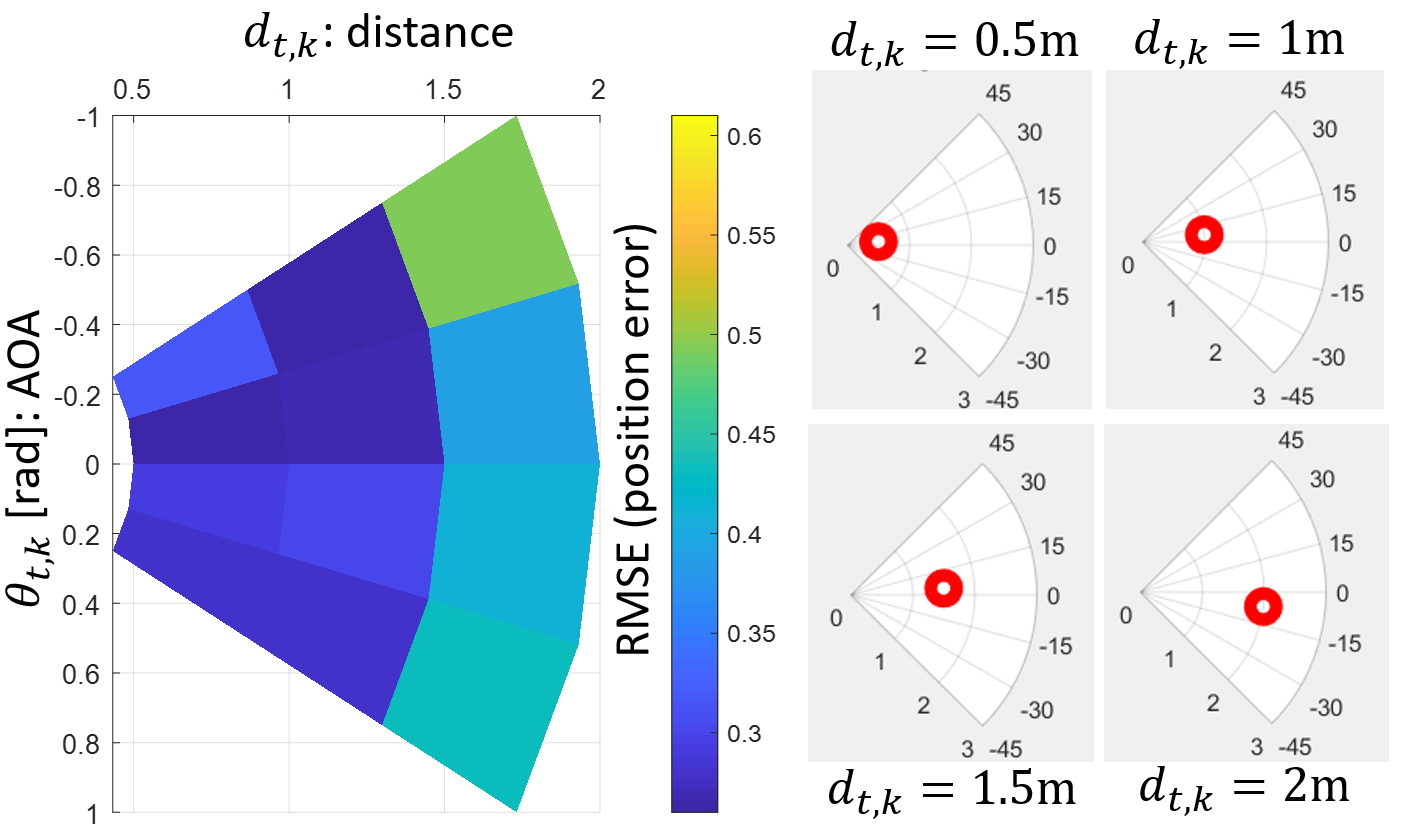} 
	\protect\caption{\label{localization} Analysis of localization RMSE for varying subject
		positions around the IR array and inside the sensor FOV.}
\end{figure}

\subsection{Sensor calibration}
\label{subsec:Sensor-calibration}
Each IoT device samples its IR array every $\triangle t=0.3$ s. Raw data are 
then forwarded to an intermediate access point via the Thread
(IEEE 802.15.4) protocol and sent to a server by exploiting the Message
Queuing Telemetry Transport (MQTT) one. The server is in charge of
data storage, caching and processing, and acts as a broker accepting
subscriptions from an ad-hoc application optimized for visualization
of the subject locations in polar or Cartesian coordinates. During
initialization, all deployed devices estimate the background
parameters $\boldsymbol{\mathbf{\mu}}_{k}$ and $\mathbf{C}_{k}$
in (\ref{eq:likelihoods}) for each location/region $k$. Re-estimation
of the parameters $\boldsymbol{\mu}_{k}$ and $\mathbf{C}_{k}$ is
required to possibly track time-varying background temperature. It
is implemented by a Multivariate Exponentially Weighted Moving Average
(MEWMA) and Covariance Matrix (MEWMC) \cite{mewma}, with smoothing
constants set to $0.99$ and $0.995$ for the $\boldsymbol{\mathbf{\mu}}_{k}$
and $\mathbf{C}_{k}$ updates, respectively. These are optimized in
\cite{thermal3} to track small background changes caused by external
thermal sources (\emph{e.g.}, air conditioners, heaters and radiators).

Locations $k$, or access areas, are pre-configured for each sensor
(wall- or ceiling-mounted) at installation time. Once the locations
$k$ are assigned, the thermal features $\mathbf{h}_{k}$ are computed
according to (\ref{model_thermal_sign}) and (\ref{model-1}). They
consist, for all cases, of the binary functions $\mathbf{b}_{k}$
and the corresponding temperature increases $\sigma\sim\mathcal{N}(\overline{\sigma},\sigma_{T}^{2})$,
whose parameters are defined in Tab. \ref{table-parameters} according
to the approximation (\ref{eq:rect}). Binary functions depend on
the number of areas $K$ to be monitored, and therefore on the sensor
layout, so they do not require online training/calibration. Considering
ceiling-mounted devices, the observed scene is divided up into $K=12$
ROIs that form a regular grid of $0.5$ m. One individual device can
monitor a $2.5$ sqm area when mounted on a $3.0$ m ceiling (faced
down). For wall-mounted devices, the monitored area is divided into
$K=5$ ROIs representing different access areas for the subjects with
corresponding AOAs defined as $\overline{\theta}_{k}=\left\{ -30^{\circ},18^{\circ},0^{\circ},18^{\circ},30^{\circ}\right\} $.
In all the proposed scenarios, up to $\zeta=3$ bodies might be co-present
within the sensor FOV.

Temperature screening adopts the parameters in Tab. \ref{table-parameters}.
Notice that the $\beta_{0}$ and $\beta_{1}$ terms depend on the
absolute temperature measurement errors and need a re-calibration
stage at device start-up. On the contrary, $\alpha_{0},\;\alpha_{1}$
(and $\alpha_{2}$ for the quadratic approximation) do not require
re-calibration, as they depend on the geometry of the acquisition
(\emph{i.e.}, the physical distance of the subject).

\begin{figure}[!t]
	\center\includegraphics[scale=0.5]{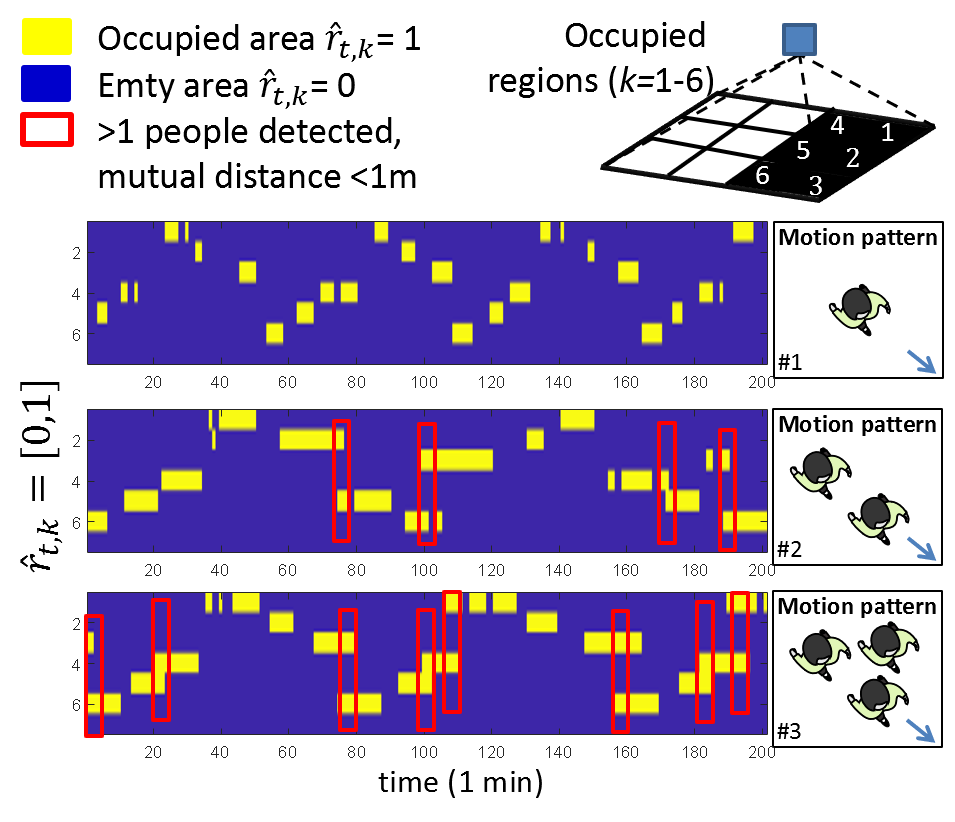} 
	\protect\caption{\label{counting and distancing} People counting and distancing alerts
		(red boxes) for ceiling-mounted devices: examples from 3 motion patterns
		featuring $\zeta=1$ (no alerts) up to $\zeta=3$ subjects.}
\end{figure}

\subsection{Localization and distancing monitoring}
\label{subsec:Localization-and-distancing}

In Tab. \ref{tab_distance_AOA}, we analyze the performance of distance
and AOA estimators, considering a wall-mounted array. Accuracy is
given in terms of Root-Mean Squared Error (RMSE). For localization
performance verification, we first consider the industrial scenario
$\#1$ where we deployed labeled landmarks in selected positions that
are occupied by the subject(s) when moving in the area. Estimator
errors are thus compared with the true positions from landmarks. The
subject is located at (true) distances from the IR array ranging from
$d_{t,k}=0.5$ m to $d_{t,k}=2.5$ m and covering access areas with
AOAs in the interval $-30^{\circ}\leq\theta_{t,k}\leq30^{\circ}$.

AOA and distance resolution tradeoff can be controlled by tuning the
model parameter $\text{\ensuremath{\overline{\sigma}_{0}}}$: large
$\text{\ensuremath{\overline{\sigma}_{0}}}$ values allow to better
track subject AOAs, but give incorrect distance estimates, generally
larger than the true ones. Smaller $\text{\ensuremath{\overline{\sigma}_{0}}}$
values produce the opposite effect. For subject positioning, we thus
choose $\text{\ensuremath{\overline{\sigma}_{0}=4.5}}$ °C, while the
other parameters are summarized in Tab. \ref{table-parameters}. The
maximum detectable body distance, above which body-induced thermal
signatures are almost indistinguishable from the (time-varying) background
could be reasonably assumed as $2.5$ m. 

In Fig. \ref{localization},
the RMSE of the location estimate, namely ($\widehat{d}_{t,k},\widehat{\theta}_{t,k}$)
is now evaluated as a function of the true target position within
the sensor FOV. In all cases, distance estimation error reduces when
the subject approaches the sensor, while the situation is reversed
for AOA estimation, namely AOA estimation improves when the subject
is entering in the sensor FOV. With respect to positioning performance,
the IR array thus behaves like a conventional radio-frequency radar
as it is sensitive in the range-azimuth domain. Notice that, although
not addressed in this paper, subject speed and motion directions could
be easily inferred by analyzing the estimated motion pattern over
the time domain.

Subject counting and physical/social distancing monitoring are now
considered in Fig. \ref{counting and distancing}, using ceiling-mounted
devices deployed as shown in the scenario $\#2$ of Fig.~\ref{layout}.
In particular, $4$ devices monitor a corridor of $10\times3$
m: cumulative alerts can be triggered, \emph{e.g.}, every minute/quarter/hour,
when the subject mutual distance falls below a threshold (set here
to $1$ m), corresponding to two distinct subjects moving in adjacent
spots. Alerts give a finer-grained information about space usage and
indicate potential criticalities depending on the specific application.
Fig. \ref{counting and distancing} highlights the estimated occupancy
patterns $\widehat{\mathbf{r}}_{t}$ (\ref{eq:map_estimate}) versus
time considering $6$ selected ROIs, with known 2D location footprint
$\mathbf{\overline{x}}_{k=1,...,6}$. Example motion patterns involve
$1$ up to $3$ people: subjects moving in adjacent ROIs cause an
alert, \emph{i.e.}, a violation of the distancing requirements, highlighted
by red boxes. By labeling a (true) violation of distancing limits
as true positive, and considering a single IR sensor array, we observe
a high precision $99$\%, corresponding to negligible false positives,
and a recall of $90$\%, namely the $90$\% of all true violations
are correctly identified. Precision is however more critical considering
that alerts are typically cumulative, as issued on minute/quarter/hour
time-frames. The combined use of multiple IR arrays is also recommended
to increase the coverage area.

As previously described, scenarios $\#1$ and $\#2$ cover almost
all situations where localization and tracking of people are critical,
such as in assisted living, homecare and smart spaces applications.
Sensors deployed as in case $\#1$ can be configured to monitor specific
locations (\emph{e.g.}, corridors, aisles and access points) by acting as virtual
fences not to be trespassed. A single wall-mounted device is sufficient
to cover an area with maximum range distance of $2.5$ m and AOAs
in the interval $-30^{\circ}\leq\theta_{t,k}\leq30^{\circ}$. In addition,
people intentions (\emph{e.g.}, motion directions and speed) and safety
information (\emph{e.g.}, distance from dangerous positions) can be easily
inferred. Unlike case $\#1$, in case $\#2$ the devices track people
density in small/medium size rooms. In such cases, the ceiling mounted
IR arrays should be installed so that the corresponding monitored
areas overlap (\emph{i.e.}, $\sim0.4\div0,5$ device/sqm density for
devices with 3-meter-high rooms). It is worth mentioning that combined 
wall- and ceiling-mounted devices deployments provide a finer-grained 
control of the space utilization. These configurations are useful 
for applications that require the simultaneous distancing monitoring 
and temperature screening, as detailed in the next Sect. \ref{subsec:Body-temperature-detection}.

Localization RMSE performance (Table \ref{tab_distance_AOA}) is in-line
with Bluetooth Low Energy (BLE) based proximity monitoring systems that are based on Received Signal Strength Indicator (RSSI) processing. In smartphone-based contact tracing apps, RSSI values are used to extract relative location estimations, with typical RMSE figures in the range
$0.5-1$ m in the short range \cite{tracing}. Notice that the proposed
setup does not need the subject to wear or possess any device; in
addition it does not pose any privacy issue as no identity tags are
tracked.

With respect to previous works on motion detection through
thermal imaging, modified K-NN and decision tree based (C4.5) models
reported precision and recall figures in the range of $80\%\div90\%$
\cite{thermal2,thermal3,thermal4,hvac} for subject counting ($1\div 4$ people
up to a distance of $3$ m). Such figures have been improved (by $5\%\div 7\%$)
in \cite{class}. However, these early works did not considered the
specific problem of localization and physical distancing monitoring.
More recent implementations that use thermal sensor with $M=768$
detectors ($32\times24$ pixels) achieve an occupancy estimation accuracy
of $98\%$ using a modified AdaBoost classifier \cite{naser_new} and considering both ceiling and wall mounted setups.

\subsection{Temperature screening: IR and radar sensor fusion}
\label{subsec:Body-temperature-detection}

For temperature screening tests, we collected measurements from $3$
different subjects ($2$ male, $1$ female) in healthy condition,
namely having $T_{\text{body}}<T_{\text{max}}$. An artificial
object with size comparable to the subject (\emph{i.e.}, head) is
also adopted to simulate Covid-19 true positives with $T_{\text{body}}>T_{\text{max}}$.
During the tests, bodies can freely move around the IR sensor array
with distance ranging between $0.1$ m, up to $1.1$ m. Besides the
IR sensor array, a low-cost radar \cite{multisensory} working in
the mmWave band was co-located (Fig. \ref{layout}) with the IR sensor
array, and used to refine body distance measurements during the screening
process. The goal is to obtain a finer-grained estimation of the body
location, still preserving the user privacy with respect to camera-based
techniques \cite{temperature4,sensors1}. The radar is equipped with
Multiple-Input-Multiple Output (MIMO) antennas ($2$ transmitters,
$4$ receivers) and implements a Frequency Modulated Continuous Wave
(FMCW) system. It works in the $77-81$ GHz band and allows for precise
range (\emph{i.e.}, distance) estimation $\widehat{d}_{t,k}$. The
subject positioning algorithm fuses the location estimates produced
by the IR sensor array with the ones obtained from the FMCW radar.
Data fusion, highlighted in Fig. \ref{block_diag}, is built sequentially
(boosting method): the location estimate is first obtained by the
Bayesian tool presented in Sect. \ref{sec:Bayesian-filtering-for}
(Pseudocode \ref{distAOA}), then cross-verified using the radar outputs
and corrected, in case of inconsistencies. Combined IR and radar-based
positioning gives a localization RMSE of $0.1$ m in the short range
($d_{t,k}\leq1$ m).

\begin{figure}[!t]
	\center\includegraphics[scale=0.54]{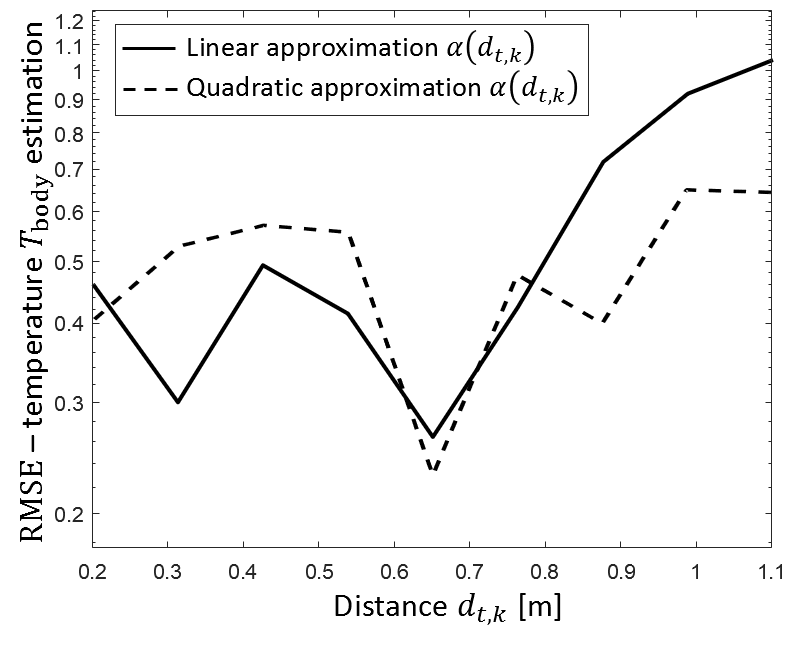} 
	\protect\caption{\label{rmse_temp} Body temperature estimation RMSE vs. subject distance
		using the IR sensor array and the mmWave radar ($0.1$ m positioning
		RMSE at $T_{\text{amb}}=23$ °C).
		Both linear and quadratic approximations are shown.}
\end{figure}

\begin{figure}[!t]
	\center\includegraphics[scale=0.6]{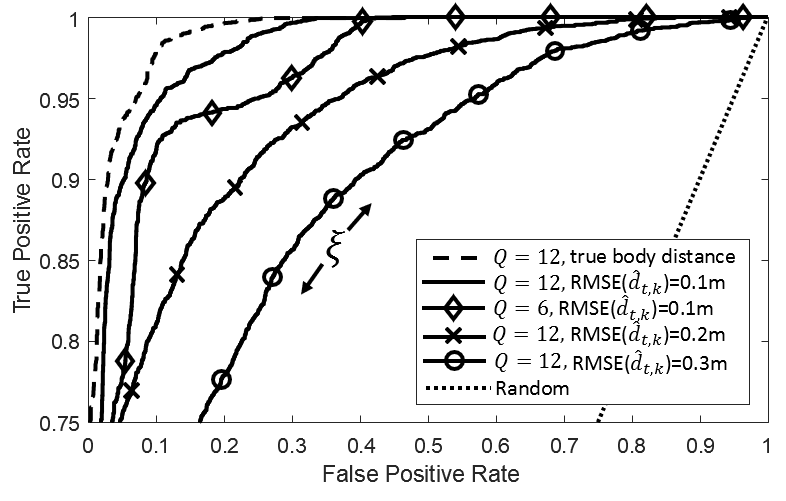} 
	\protect\caption{\label{roc}ROC curve for temperature screening (varying number of
		samples $Q$ and distance estimation accuracies).}
\end{figure}

Using positioning refinements, temperature estimation is obtained
by solving (\ref{eq:linear}). Estimation RMSE is analyzed in Fig.
\ref{rmse_temp} for varying subject distances, and $T_{\mathrm{amb}}=23\lyxmathsym{\textdegree}\mathrm{C}$.
Linear and quadratic approximations for $\alpha(d_{t,k})$ in (\ref{eq:alf})
are also compared. The linearized model gives a body temperature estimate
with average RMSE of $0.37\lyxmathsym{\textdegree}\mathrm{C}$ for
distances lower than $0.75$ m. For larger distances (but still $\leq1.1$
m), the quadratic approximation performs better, although the RMSE
increases up to $0.45\,\lyxmathsym{\textdegree}\mathrm{C}$. The maximum
body distance for screening could be thus reasonably set to $1$ m.

In Fig. \ref{roc}, we analyze the Receiver Operating Characteristic
(ROC) figures for varying number of samples ($Q=6,12$) and positioning
accuracy, with body location RMSE ranging from $0.10$ m to $0.32$
m. The ROC curve using true body distance information is also superimposed
in dashed line as benchmark while the random (\emph{i.e.}, trivial)
detector is shown as well. In particular, we consider $2888$ true
negative examples, namely healthy subjects with $T_{\mathrm{body}}$
ranging from $36$ \lyxmathsym{\textdegree}C to $37.2$ \lyxmathsym{\textdegree}C,
and $1800$ true positives, with $T_{\mathrm{body}}$ ranging from 
$37.8$ \lyxmathsym{\textdegree}C to $38.5$ \lyxmathsym{\textdegree}C.
True temperatures are measured by a contact thermometer. Pre-screening inspection 
systems \cite{sensors1} need high true positive rates (TPR), or recall, to
maximize the number of subjects correctly detected as unhealthy. In
addition, false positive rates (FPR) should be kept as low as possible
to limit post-screening operations. Considering $Q=12$ samples, corresponding
to an inspection interval of $\sim4$ seconds, the TPR and the FPR
are $97\%$ and $15\%$, respectively, for the detector threshold
set to $\xi=-0.2$ (optimized for maximum recall), and subject 
positioning RMSE of $0.1$ m. Distancing
errors have larger impact on screening performance, compared to inspection
interval $Q$. In fact, reducing the inspection interval to $\sim2$
seconds ($Q=6$) brings down the TPR to $94$\% and increases the
FPR to $18\%$. Positioning errors (RMSE $0.2$ m) are still acceptable
but reduce the TPR to $92\%$ and and increase the FPR to $25\%$.
Screening performance experiences further degradation for larger RMSE.

\begin{table}[tp]
	\begin{centering} \protect\caption{\label{confusion} Temperature screening for $Q=12$ and $\xi=-0.2$: precision and recall figures using IR sensor and data fusion with radar for distancing refinement.}
		\begin{tabular}{l|c|c|c|c|}
			\cline{2-5} 
			& \multicolumn{2}{c|}{IR sensor only} & \multicolumn{2}{l|}{Radar + IR sensor}\tabularnewline
			& \multicolumn{2}{c|}{(RMSE $0.32$ m)} & \multicolumn{2}{c|}{(RMSE $0.10$ m)}\tabularnewline
			\cline{2-5} 
			& Recall & Precision & Recall & Precision\tabularnewline
			\hline 
			\multicolumn{1}{|l|}{$\mathrm{F}_{1}:T_{\mathrm{body}}>T_{\mathrm{max}}$} & \multicolumn{1}{c|}{$88.0\%$} & \multicolumn{1}{c|}{$79.1\%$} & $97.0\%$ & $87.5\%$\tabularnewline
			\hline 
			\multicolumn{1}{|l|}{$\mathrm{F}_{0}:T_{\mathrm{body}}<T_{\mathrm{max}}$} & \multicolumn{1}{c|}{$84.0\%$} & \multicolumn{1}{c|}{$88.1\%$} & $90.4\%$ & $98.0\%$\tabularnewline
			\hline 
		\end{tabular}
		\par\end{centering}
	\medskip{}
\end{table}

Considering the same examples, precision and recall figures are now
compared in Tab. \ref{confusion} using both radar and IR array for
distance monitoring. The wall-mounted IR-based devices configured for localization
gives an RMSE of $0.32$ m (as in Tab. \ref{tab_distance_AOA}) while
screening precision and recall are $79.1\%$ and $88\%$, respectively.
The combined use of the IR array and the radar gives an RMSE of $0.1$
m and thus improve both figures that are now $87.5\%$ and $97\%$,
respectively. These results are in-line with RGB-thermal image-based
screening tools with SVM classification \cite{sensors1} that report a 
recall figure equal to  $85.7\%$ and a specificity, namely $1-\mathrm{FPR},$ 
equal to the $90.1\%$ one. In the same reference \cite{sensors1}, conventional 
fever screening methods based on electronic thermometers show recall and 
specificity values equal to $60.7\%$ and $86.4\%$, respectively.

\section{Conclusions}
\label{sec:conc} The paper proposed a Bayesian tool for joint body
localization and temperature screening that is based on the real-time
analysis of infrared (IR) body emissions using devices equipped with low-cost 
thermopile-type passive IR sensor arrays. A statistical model is validated 
experimentally in an industrial test plant to represent the thermal signatures 
induced by bodies at different locations. The model is general enough to support
different device layouts (ceiling- and wall-mounted devices) and it
has been verified experimentally targeting the passive detection of
body motion directions, distances from the sensor and mutual body
distances, to monitor crowded areas. Thus, the proposed sensing platform
can be easily adapted for applications ranging from assisted living
to homecare, and smart spaces. Contactless temperature pre-screening,
rooted at Bayesian decision theory, is integrated with body localization
that fuses IR emissions and backscattered radar signals for finer-grained
positioning. It thus allows non-invasive body diagnosis by letting
the subject freely move in the surroundings of the IR sensor, during
the measurement process. Screening is designed to minimize false negatives
with performances that are in-line with state of the art inspection
tools, but, preserving subject privacy by avoiding imaging from video
cameras.

\section*{Appendix: Envisense system description}
The infra-red array sensor is supported by the multi-sensor box Envisense Smart equipped with a 2.4 GHz IEEE 802.15.4 radio interface. The infra-red array performs $10$ measurements per second of the thermal profile of the space in front of itself, with a range of coverage that reaches up to $7$ m and an angular opening of $60$ deg. Sensors are identified by a unique 64-bit number corresponding to the address of their IEEE 802.15.4 radio module. They transmit infra-red array data with a low-power 2.4 GHz mesh network, each 8x8 thermal map sampled at 10 Hz is sent to the data collector called EnviCore. The thermal map sample is quantized with 8-bits representing the temperature ranging from 0 to 64 °C with a 0.25 °C resolution. Data collected by the gateway is converted to a JSON data packet, marked with a absolute timestamp in order to keep the timing information and allow the synchronization with other sensors data. Finally, all collected and converted data is sent to the processing server by a Wi-Fi connection and MQTT protocol.
In order to have the platform working the following configurations steps shall be done: i) EnviSense configuration using the Android App and a smartphone with NFC support; ii) EnviCore configuration with its web pages; iii) processing server configuration.

In order to configure the Envisense device an Android App is provided. The Android Smartphone used for configuration must have NFC technology support in order to communicate with the EnviSense. After the app is launched the configuration is read by simply moving close the smartphone to the device between the solar panel and the PIR. As soon as the configuration is read all sensors are enabled. See \cite{envisense} for further technical information.

\end{document}